\documentclass[a4paper,fleqn,usenatbib]{mnras}

\usepackage[T1]{fontenc}
\usepackage{ae,aecompl}
\usepackage{graphicx}
\usepackage{amsmath}
\usepackage{amssymb}
\usepackage{subcaption}
\usepackage{bm}
\usepackage{newtxtext,newtxmath}
\usepackage{enumitem}
\usepackage{lastpage}
\usepackage{CJK}
\usepackage{booktabs}

\usepackage{ulem}

\usepackage{xcolor}
\usepackage{soul}

\begin{document}
\begin{CJK}{UTF8}{gbsn}
\normalem

\newcommand{\HI}{H{\sc ~i}\ }
\newcommand{\HII}{H{\sc ~ii}\ }
\newcommand{\HeI}{He{\sc ~i}\ }
\newcommand{\HeII}{He{\sc ~ii}\ }
\newcommand{\HeIII}{He{\sc ~iii}\ }
\newcommand{\Msun}{{M_{\odot}}\ }
\newcommand{\CIV}{C{\sc ~iv}\ }
\newcommand{\NV}{N{\sc ~v}\ }
\newcommand{\OVI}{O{\sc ~vi}\ }
\newcommand{\OVII}{O{\sc ~vii}\ }
\newcommand{\SiIV}{Si{\sc ~iv}\ }

\definecolor{darkgreen}{RGB}{0,100,0}

\def\gsim{\;\rlap{\lower 2.5pt
 \hbox{$\sim$}}\raise 1.5pt\hbox{$>$}\;}
\def\lsim{\;\rlap{\lower 2.5pt
   \hbox{$\sim$}}\raise 1.5pt\hbox{$<$}\;}

\title[Simulations of radiative turbulent mixing layers]{Simulations of radiative turbulent mixing layers}
\author[Ji, Oh \& Masterson]{Suoqing Ji (季索清),$^{1,2}$\thanks{Email: suoqing@caltech.edu} S. Peng Oh$^{1}$ and Phillip Masterson$^{1}$ \\
\\
$^{1}$ Department of Physics, University of California, Santa Barbara, CA 93106, USA.\\
$^{2}$ TAPIR, Walter Burke Institute for Theoretical Physics, California Institute of Technology, Pasadena, CA 91125, USA.}

\setlength{\topmargin}{-0.6cm}
\date{Accepted 0000. Received 0000; in original form 0000}

\pagerange{\pageref{firstpage}--\pageref{LastPage}}
\pubyear{2018}

\label{firstpage}
\maketitle

\begin{abstract}

Radiative turbulent mixing layers should be ubiquitous in multi-phase gas with shear flow. They are a potentially attractive explanation for the high ions such as \OVI seen in high velocity clouds and the circumgalactic medium (CGM) of galaxies. We perform 3D MHD simulations with non-equilibrium (NEI) and photoionization modeling, with an eye towards testing simple analytic models. Even purely hydrodynamic collisional ionization equilibrium (CIE) calculations have column densities much lower than observations. Characteristic inflow and turbulent velocities are much less than the shear velocity, and the layer width $h \propto t_\mathrm{cool}^{1/2}$ rather than $h \propto t_\mathrm{cool}$. Column densities are not independent of density or metallicity as analytic scalings predict, and show surprisingly weak dependence on shear velocity and density contrast. Radiative cooling, rather than Kelvin-Helmholtz instability, appears paramount in determining the saturated state. Low pressure due to fast cooling both seeds turbulence and sets the entrainment rate of hot gas, whose enthalpy flux, along with turbulent dissipation, energizes the layer. Regardless of initial geometry, magnetic fields are amplified and stabilize the mixing layer via magnetic tension, producing almost laminar flow and depressing column densities. NEI effects can boost column densities by factors of a few. Suppression of cooling by NEI or photoionization can in principle also increase \OVI column densities, but in practice is unimportant for CGM conditions. To explain observations, sightlines must pierce hundreds or thousands of mixing layers, which may be plausible if the CGM exists as a ``fog'' of tiny cloudlets. 
\end{abstract}


\section{Introduction}
\label{section:intro}

In recent years, particularly since the launch of the Cosmic Origins Spectrograph (COS) on the Hubble Space Telescope (HST), studies of the circumgalactic medium (CGM) in the hot halos surrounding galaxies have received fresh impetus and significantly better observational constraints (for a recent comprehensive review, see \citealt{tumlinson17}). The CGM is an exciting new frontier in studies of galaxy formation, and introduces a new set of interesting puzzles and constraints. Chief amongst these is the ubiquity of high ions, most notably \OVI, detected in absorption in quasar sightlines intersecting the CGM (e.g., \citealt{tumlinson2005hot, prochaska11, tumlinson11, stocke2013characterizing, savage2014properties, werk14, lehner15, johnson15, burchett2016deep, prochaska17, keeney2017characterizing}). Such observations indicate a significant mass reservoir in \OVI gas, on the order of $\sim 10^{9}-10^{10} \, M_{\odot}$. Such high ions are also seen in absorption spectra of high velocity clouds (HVCs) around the Milky Way (e.g, \citealt{savage14}). The origin of such gas is controversial, with line ratios and profiles sometimes supporting either a photoionized (e.g., \citealt{muzahid15}) or collisionally ionized (e.g., \citealt{heckman02,bordoloi16}) origin, at times even within the same multi-component absorber. The truth is likely to be complex, with both mechanisms playing an important role. 

The case for either mechanism is not without difficulties. A frequent objection against photoionization models is the large path lengths ($\gtrsim 100\,$kpc) required to obtain the observed \OVI columns and \NV/\OVI column density ratios \citep{werk16}. This could potentially be resolved by a thick low pressure shell at the halo outskirts \citep{stern2018does}. However, the troubles faced by collisional ionization models are even more daunting (e.g., \citealt{faerman17, mcquinn18}). For instance, in collisional ionization equilibrium, \OVI is abundant only over a very narrow range in temperature, peaking at $\sim 3 \times 10^5 \ \mathrm{K}$ with significant abundance out to $10^6 \ \mathrm{K}$. This is quite different from the virial temperatures of the halos in which it is detected, suggesting relatively little gas is in this phase, as indeed is found in cosmological simulations \citep{liang16}. Moreover, if \OVI gas is associated with virialized gas, there should be a steep dependence of \OVI abundance with halo virial temperature (which, while not directly observable, varies with stellar mass).  On the contrary, observed \OVI column densities are relatively independent of stellar (and thereby halo) mass. Most importantly, this temperature corresponds to the exact peak of the cooling curve, and gas abundant in \OVI should quickly cool out of this phase. Thus, a ``cooling flow'' interpretation of \OVI observations requires an enormous mass flux of gas cooling through \OVI temperatures --- roughly $\sim 30 \, M_{\odot} \, \mathrm{yr^{-1}}$ \citep{mcquinn18}, which is much higher than the star formation rates found in the host galaxies. Such a scenario would only make sense if there is a way to rapidly recycle cold gas back to higher temperatures, so that the actual amount of cold gas in steady state is relatively small. 

This work focuses on one of the potentially promising \OVI sources -- turbulent mixing layers (TMLs; \citealt{begelman90, slavin93}). TMLs are part of a broader class of models which utilize the fact that the interface between hot and cold gas is unlikely to be a sharp contact discontinuity. Any diffusive heat transport process, such as thermal conduction \citep{borkowski90, gnat10}, collisionless cosmic-ray scattering \citep{wiener17-cold-clouds}, or turbulent mixing, should broaden the transition and produce intermediate temperature gas. TMLs arise due to turbulence at the interfaces between gas of different phases (e.g., driven by Kelvin Helmholtz instabilities, background turbulence in the hot medium, radiative shocks around SNRs or wind-driven bubbles, or cloud crushing due to shock passage over a gas cloud), which drives mixing. In hydrodynamic mixing, the hot mass flux per unit area, $\dot{m}_\mathrm{hot} \sim \eta_{\rm h} \rho_\mathrm{hot} v_\mathrm{t}$, and the cold mass flux $\dot{m}_\mathrm{cold} \sim \eta_{\rm c} \rho_\mathrm{cold} l/t_\mathrm{KH} \sim (\rho_\mathrm{hot}\rho_\mathrm{cold})^{1/2} v_\mathrm{t}$, where $v_\mathrm{t}$ is the turbulent velocity, $l$ is the outer scale of turbulence, and $t_\mathrm{KH} \sim (\rho_\mathrm{hot}/\rho_\mathrm{cold})^{1/2} l/v_\mathrm{t}$, mix to produce gas with mean temperature \citep{begelman90}:
\begin{align}
\bar{T}_{\rm mix} &\approx \frac{\dot{m}_\mathrm{hot} T_\mathrm{hot} + \dot{m}_\mathrm{cold} T_\mathrm{cold}}{(\dot{m}_\mathrm{hot} +  \dot{m}_\mathrm{hot} )} \nonumber \\ 
&\approx \frac{[\eta_{\rm h} + \eta_{\rm c}(T_{\rm c}/T_{\rm h})^{1/2}]}{[\eta_{c} + \eta_{\rm h}(T_{\rm c}/T_{\rm h})^{1/2}]} (T_\mathrm{hot} T_\mathrm{cold})^{1/2} \nonumber \\ &\equiv \zeta (T_\mathrm{hot} T_\mathrm{cold})^{1/2}, 
\end{align}
i.e. roughly the geometric mean of the two temperatures. For cold gas clouds with $T\sim 10^{4}$K moving through host halos with virial temperature gas $T \sim 10^{6}$K, this give mixed gas of $T \sim 10^{5}$K, precisely the intermediate temperature range we are seeking. 
In steady state, if radiative cooling balances the energy flux of the entrained hot gas, i.e., $l/v_\mathrm{t} \sim t_\mathrm{cool}$, then the thickness of the mixing layer is $l \sim v_\mathrm{t} t_\mathrm{cool}$, which similar to radiative shocks and hot winds, produces an ion column density \citep{heckman02, mcquinn18}: 
\begin{equation}
N_\mathrm{OVI} \sim \eta f_\mathrm{OVI} \left[ f_\mathrm{O} \right]_{\odot} Z n v_\mathrm{t} t_\mathrm{cool} \sim 8 \times 10^{12} \ \mathrm{cm^{-2}} \eta_{0.1} v_{2}
\label{eq:TML_column}
\end{equation}
where $v_{2} \equiv (v/100 \, \mathrm{km \, s^{-1}})$, $\eta_{h,0.1} \equiv \eta_{\rm h}/0.1$,  $t_\mathrm{cool}$ is evaluated at the peak \OVI abundance $f_\mathrm{OVI} = 0.2$ at $T=10^{5.5} \, \mathrm{K}$, and $\eta$ encodes various unknowns, such as entrainment efficiency and geometric factors . We assume $\left[ f_\mathrm{O} \right]_{\odot} = 4.56 \times 10^{-4}$ \citep{asplund09}. A more careful calculation which follows ionization fractions over the cooling lifetime of the gas parcel can further reduce the above estimate by up to an order of magnitude \citep{slavin93}.

TMLs thus have a number of attractive features.  Since $t_\mathrm{cool} \propto 1/(n \, Z)$ in the crucial temperature range, in CIE the metal column density above is approximately independent of the underlying density and metallicity. TMLs potentially obviate the need for a massive cooling flow, as gas continually cycles between cold and hot phases. Moreover, they explain a number of observational features: the coincidence in velocity space between low and high ions in many (though not all) systems \citep{werk14,werk16}, and an observed correlation between column density and b-parameter, as predicted by Eq. \eqref{eq:TML_column}. However, the predicted columns are 10 -- 100 times lower than in observations, thus requiring that a sightline pierce 10 -- 100 such interfaces in order to explain the observations.

Besides early analytic work, TMLs have also been the subject of 2D hydrodynamic simulations, most notably in a series of papers by K. Kwak and collaborators \citep{kwak10, kwak11, henley12, kwak15}. However, they did not compare their simulation results to analytic models. Since both mixing and radiative cooling can occur on shorter timescales than ionization or recombination, they also incorporate non-equilibrium ionization (NEI) in their simulations, and find that NEI effects boost high ion abundances and column densities (\CIV, \NV, \OVI) by a factor of a few, compared to collisional ionization equilibrium (CIE) calculations. Ion line ratios (\CIV/\NV, \NV/\OVI) were in rough agreement with observations. TMLs were also simulated in 3D MHD CIE calculations by \citet{esquivel06}. However, the simulations were not run for long enough for effective mixing (or a stable equilibrium) to develop, so it is difficult to draw firm conclusions. 

In this paper, we seek to improve our understanding of TMLs on several fronts. Firstly, we compare our hydrodynamic simulations to existing analytic models, an exercise which surprisingly has not been carried at in the literature. We find surprising disagreement with expected scalings. Secondly, we study the impact of photoionization. Photoionization and collisional ionization are frequently treated in isolation; most studies of TMLs consider collisional ionization exclusively with the exception of a semi-analytic model which includes photoionization \citep{slavin93}, while many observational studies focus on CLOUDY photoionization models. In reality, of course, they operate concurrently. Photonionization breaks the density independence of TMLs, since the ionization parameter depends on density. It changes ion ratios, boosting the relative abundance of low ions. Photoionization also changes the balance between mixing and cooling by changing the cooling function and reducing the efficacy of radiative cooling (since ions are over-ionized for a given temperature). Thirdly, we run 3D MHD simulations to consider the effect of magnetic fields on TMLs, taking care to ensure that the simulations reach a steady state. magnetic fields can be amplified by and substantially change the character of turbulence. MHD forces introduce anisotropy and can suppress mixing, but these effects have not been quantified in the context of TMLs. Finally, we reconsider NEI effects in light of the above considerations. 

The outline of this paper is as follows. In \S\ref{sect:methods}, we describe our simulation method. In \S\ref{sect:results}, we present our results, comparing hydrodynamic simulations to analytic models, contrasting hydrodynamic vs. MHD simulations with varying field geometry, the impact of photoionization and NEI, and convergence tests. We conclude in \S\ref{sect:conclusions}.

\section{Methods}
\label{sect:methods}

  We use the FLASH code \citep{fryxell00} developed by the FLASH Center of the University of Chicago for our simulations, where a directionally unsplit staggered mesh (USM) MHD solver is adopted to solve the equations of inviscid ideal magnetohydrodynamics:
  \begin{subequations}
    \begin{gather}
      \frac{\partial\rho}{\partial t} + \bm{\nabla} \cdot (\rho \bm {v}) = 0 \\
      \frac{\partial\rho \bm{v}}{\partial t} + \bm{\nabla} \cdot (\rho \bm{v} \bm{v} - \bm{B} \bm{B}) + \bm{\nabla} p_* = 0 \\
      \rho T \left(\frac{\partial s}{\partial t} + \bm {v} \cdot \nabla s \right) = - \mathcal{L} \\
      \frac{\partial \bm{B}}{\partial t} + \bm{\nabla} \cdot (\bm{v}\bm{B} - \bm{B} \bm{v}) = 0 \\
      \nabla \cdot \bm{B} = 0
    \end{gather}
    \label{eq:MHD}
  \end{subequations}
  where $p_* = p + B^2 / (8 \pi)$ is the total pressure including both gas pressure $p$ and magnetic pressure $B^2 / (8 \pi)$, $s = c_\mathrm{p} \mathrm{ln} (P \rho^{-\gamma})$ is entropy per unit mass,, and $\mathcal{L}$ is the cooling rate. The MHD solver is based on a finite-volume, high-order Godunov method combined with a constrained transport (CT) type of scheme which ensures the solenoidal constraint of the magnetic fields on a staggered mesh geometry \citep{tzeferacosetal12, lee2013solution}.

  To set the simulations up, firstly we initiate a 3D domain with $100 \ \mathrm{pc}$ in width and $300 \ \mathrm{pc}$ in height, with the resolution of $128\times128\times364$ as standard runs. Our coordinate system is: $x$ is the flow direction, $z$ is normal to the cold/hot interface, and $y$ is the third dimension. The resolution is not excessively high, since we evolve a large number of ions, which makes the simulations both time consuming and memory intensive. Outflow boundary conditions are applied to top and bottom boundaries, and periodic boundary conditions to the sides. Cold and and hot gas with a velocity shear are filled in the top ($z > 0 \ \mathrm{pc}$) and bottom ($z < 0 \ \mathrm{pc}$) regions of the domain respectively, with the top (cold) half initially stationary and the bottom (hot) half moving at velocity $v_\mathrm{hot}$. The fluid quantities of these two phases are connected smoothly with a $\mathrm{tanh}$ function:
  \begin{align}
    \rho(z) &= \rho_\mathrm{hot} + \frac{\rho_\mathrm{top} - \rho_\mathrm{hot}}{2}
     \left[1+\mathrm{tanh}\left(\frac{z}{a}\right) \right] \\
    v_x(z) &= \frac{v_\mathrm{bot}}{2}
     \left[1-\mathrm{tanh}\left(\frac{z}{a}\right) \right],
  \end{align}
  where $\rho$ stands for density, $v_x$ is the fluid velocity along $x$ direction, and $a=2.5 \ \mathrm{pc}$ the half-length of the shear layer. This sets up a two-phase medium with cold, dense gas at the top and hot, diffuse gas on the bottom. To excite a Kelvin-Helmholtz instability at the interface of the two phases, we initialize a velocity perturbation along the $z$ direction:
  \begin{align}
    v_z = \delta v\ \mathrm{exp} \left[-\left(\frac{z}{a}\right)^2\right] 
      \mathrm{sin}\left(\frac{2\pi x}{\lambda}\right)
      \mathrm{sin}\left(\frac{2\pi y}{\lambda}\right),
  \end{align}
  where $\lambda$ is the wave length of perturbation, and $\delta v$ is set to $v_\mathrm{bot}/100$. The parameters for fiducial simulations are listed in Table \ref{tb:fiducial}.

\begin{table*}
\renewcommand{\arraystretch}{1.5}

\begin{center}
  \begin{tabular}{cccccc}
    \toprule[1pt]\midrule[0.4pt]
     \bf Domain Size ($\mathrm{pc}$) & \bf Resolution & \bf Number Density ($\mathrm{cm^{-3}}$) & \bf Temperature ($\mathrm{K}$) & \bf Velocity Shear ($\mathrm{km/s}$) & \bf Wavelength ($\mathrm{pc}$) \\ \midrule[0.4pt]
    $100^2\times300$ 
    & $128^2\times384$
    & \begin{tabular}{@{}c}
        $n_\mathrm{cold} = 1.6\times10^{-2}$ \\
        $n_\mathrm{hot} = 1.6\times10^{-4}$
      \end{tabular}
    & \begin{tabular}{@{}c}
        $T_\mathrm{cold} = 10^4$ \\ 
        $T_\mathrm{hot} = 10^6$ 
      \end{tabular}
    & \begin{tabular}{@{}c}
        $v_x = 100$ \\ 
        ($\mathcal{M} = 0.69$)
      \end{tabular}
    & $\lambda = 100$ \\
  \bottomrule[1pt] 
  \end{tabular}
\end{center}
\caption{Parameters for fiducial simulations. The Mach number $\mathcal{M}\equiv v_x/c_\mathrm{s}$, where $c_\mathrm{s}$ refers to the sound speed in the hot medium.}
\label{tb:fiducial}
\end{table*}

\begin{table*}
\renewcommand{\arraystretch}{1.5}

\begin{center}
  \begin{tabular}{cccccccccccc}
    \toprule[1pt]\midrule[0.4pt]
       & \multicolumn{2}{c}{\bf Metallicity} & \multicolumn{5}{c}{\bf Cooling Curve Adjustment}  & \multicolumn{2}{c}{\bf NEI} & \multicolumn{2}{c}{\bf Photoionization} \\ \cmidrule[0.4pt](lr){2-3} \cmidrule[0.4pt](lr){4-8} \cmidrule[0.4pt](lr){9-10} \cmidrule[0.4pt](lr){11-12}
    \bf Value
      & $0.1Z_\odot$   & $Z_\odot$       & $1$                & $0.1$          & $10$      &  $0$     &   time-dependent           & not adopted      & adopted            & not included    & included \\
    \bf Name
      & \tt Z0.1   & \tt Z1    & (none)    & \tt cool0.1      & \tt cool10    & \tt cool0     & \tt cool-timedp       & (none) & \tt nei    & (none)      & \tt photo \\ \cmidrule[0.4pt](lr){4-7} \cmidrule[0.4pt](lr){9-10}
    \bf Note
      & \multicolumn{2}{c}{} & \multicolumn{4}{c}{a control pre-factor to adjust cooling time} &  \multicolumn{1}{c}{} &   \multicolumn{2}{c}{non-equilibrium ionization}  & \multicolumn{2}{c}{}  \\ 
  \bottomrule[1pt]
  \end{tabular}
\end{center}

\vspace{0.2cm}

\begin{center}
\begin{tabular}{cccccccc}
  \toprule[1pt]\midrule[0.4pt]
    & \multicolumn{7}{c}{\bf Magnetic Field Strength ($\mathrm{\mu G}$) and Orientation} \\ \cmidrule[0.4pt](lr){2-8}
  \bf Value 
    & $\bm{B}=0$      & $B_x=0.035$ & $B_y=0.035$ & $B_z=0.035$ & $B_x=0.35$   & $B_y= 0.35$    & $B_z=0.35$ \\
  \bf Name   
    & (none) & \tt Bx-sml  & \tt By-sml  & \tt Bz-sml  & \tt Bx-lgr     & \tt By-lgr     & \tt Bz-lgr     \\ \cmidrule[0.4pt](lr){3-5} \cmidrule[0.4pt](lr){6-8}
  \bf Note   
    &    & \multicolumn{3}{c}{$\beta=430$, $\mathcal{M}_\mathrm{A}=13$} & \multicolumn{3}{c}{$\beta=4.3$, $\mathcal{M}_\mathrm{A}=1.3$}      \\ 
  \bottomrule[1pt]
\end{tabular}
\end{center}

\vspace{0.2cm}

\begin{center}  
\begin{tabular}{cccccc}
  \toprule[1pt]\midrule[0.4pt]
    & \multicolumn{2}{c}{\bf Density Contrast} & \multicolumn{2}{c}{\bf Ambient Pressure ($\mathrm{erg/cm^3}$)}   \\ \cmidrule[0.4pt](lr){2-3} \cmidrule{4-5}
  \bf Value 
    & $100$ & $1000$ & $2.1\times10^{-14}$ & $2.1\times10^{-15}$ \\
  \bf Name 
    & (none)   & \tt contr-lgr      & (none)       & \tt pres-sml  \\
  \bf Note
    & \begin{tabular}{@{}c}
        $n_\mathrm{cold} = 1.6\times10^{-2} \ \mathrm{cm^{-3}}$ \\
        $n_\mathrm{hot} = 1.6\times10^{-4} \ \mathrm{cm^{-3}}$ \\
        $T_\mathrm{cold} = 10^4 \ \mathrm{K}$ \\ 
        $T_\mathrm{hot} = 10^6 \ \mathrm{K}$ 
      \end{tabular}
    & \begin{tabular}{@{}c}
        $n_\mathrm{cold} = 1.6\times10^{-1} \ \mathrm{cm^{-3}}$ \\
        $n_\mathrm{hot} = 1.6\times10^{-4} \ \mathrm{cm^{-3}}$ \\
        $T_\mathrm{cold} = 10^3 \ \mathrm{K}$ \\ 
        $T_\mathrm{hot} = 10^6 \ \mathrm{K}$ 
      \end{tabular}
    & \begin{tabular}{@{}c}
        $n_\mathrm{cold} = 1.6\times10^{-2} \ \mathrm{cm^{-3}}$ \\
        $n_\mathrm{hot} = 1.6\times10^{-4} \ \mathrm{cm^{-3}}$ \\
        $T_\mathrm{cold} = 10^4 \ \mathrm{K}$ \\ 
        $T_\mathrm{hot} = 10^6 \ \mathrm{K}$ 
      \end{tabular}
    & \begin{tabular}{@{}c}
        $n_\mathrm{cold} = 1.6\times10^{-3} \ \mathrm{cm^{-3}}$ \\
        $n_\mathrm{hot} = 1.6\times10^{-5} \ \mathrm{cm^{-3}}$ \\
        $T_\mathrm{cold} = 10^4 \ \mathrm{K}$ \\ 
        $T_\mathrm{hot} = 10^6 \ \mathrm{K}$ 
      \end{tabular} \\   
  \bottomrule[1pt]
\end{tabular}
\end{center}

\vspace{0.2cm}

\begin{center}
  \begin{tabular}{cccccccccc}
    \toprule[1pt]\midrule[0.4pt]
      & \multicolumn{3}{c}{\bf Velocity Shear ($\mathrm{km/s}$)} & \multicolumn{2}{c}{\bf Domain Size ($\mathrm{pc}$)}  & \multicolumn{2}{c}{\bf Wavelength ($\mathrm{pc}$)} & \multicolumn{2}{c|}{\bf Resolution} \\ \cmidrule[0.4pt](lr){2-4} \cmidrule[0.4pt](lr){5-6} \cmidrule[0.4pt](lr){7-8} \cmidrule[0.4pt](lr){9-10}
    \bf Value
      & $100$  & $50$ & $200$ & $100^2\times300$ &  $200^2\times300$ & $100$  & $50$ & $128^2\times384$  & $256^2\times768$ \\
    \bf Name
      & (none)  & \tt vel-sml & \tt{vel-lgr}  & (none) & \tt size-lgr & (none) & \tt wave-sml & (none) & \tt res-lgr \\ \cmidrule[0.4pt](lr){9-10}
    \bf Note
      & $\mathcal{M} = 0.69$ & $\mathcal{M} = 0.34$ & $\mathcal{M} = 1.37$ & \multicolumn{2}{c}{} & \multicolumn{2}{c}{} & \multicolumn{2}{c}{convergence tests} \\ 
    \bottomrule[1pt]
  \end{tabular}
\end{center}

\caption{Varying simulation parameters with their corresponding name elements, where the name elements are omitted (set to none) for the fiducial parameter values. The plasma beta $\beta\equiv P_\mathrm{gas} / P_\mathrm{mag}$, the Mach number $\mathcal{M}\equiv v_x/c_\mathrm{s}$, where $c_\mathrm{s}$ refers to the sound speed in the hot medium, and the Alfv\'en Mach number $\mathcal{M}_\mathrm{A} \equiv v_x / v_\mathrm{A}$.}
\label{tb:parameter}
\end{table*}

  Some aspects of these initial and boundary conditions are worth discussion. Simulations of jet entrainment and the Kelvin Helmholtz instability use both periodic boundary conditions and long boxes with outflow boundary conditions. The latter can capture the spatial evolution of the flow (e.g., the thickening of the mixing layer), but --- except for an extremely long box --- can not capture the saturated, non-linear steady state outcome of the Kelvin Helmholtz instability. The latter requires periodic boundary conditions, which correspond to a section of the wake. A finite periodic box of course means that only unstable modes $\lambda = L/n$ which are harmonics of the box-size  $L$ can be captured. In cases where the mixing layer width is a significant fraction of the box length $L$, we have doubled the box length to verified that finite box-size effects are not important.

  Our initial conditions are meant to mimic the flow of hot gas past a cold cloud. Since the mass of cold gas in the box is a factor of $\sim \delta$ higher, the hot gas can decelerate significantly as it transfers momentum to the cold gas. We have checked that the hot gas velocity $v_x$ at the lower boundary stays close to its initial velocity throughout the run, as it should if the box is sufficiently long in the vertical direction. We have also explicitly verified the Galilean invariance of our simulations: mixing layer properties are identical if the hot gas is initially stationary and the cold gas moves.

In order to predict column densities, we calculate non-equilibrium ionization (NEI) for $5$ elements: He, C, N, O and Si. All the ions of these elements are initialized in collisional / photoionization equilibrium and evolved with time. We adopt a fiducial metallicity of solar metallicity, although we also explore the effects of different metallicity. In the COS Halo sample, the median metallicity is $Z \approx 0.3 Z_{\odot}$ with a 95\% confidence interval ranging from $0.02 - 3 \, Z_{\odot}$ \citep{prochaska17}. Note that the mass flux into the mixing layer is dominated by the cold gas, which should be higher metallicity. We modified the FLASH NEI solver to include the contribution of photoionization, by calculating photoionization rates in the optically thin regime:
  \begin{align}
    \Gamma^\gamma = \int_{\nu_T}^{\infty} \frac{4\pi J(\nu)}{h\nu} \sigma(\nu) d\nu,
  \end{align}
  where $\Gamma^\gamma$ is the photoionization rate from the isotropic metagalactic ionizing background, ignoring any local galactic contribution, $\nu_T$ the threshold frequency for photoionization, $J(\nu)$ the background radiation intensity, and $\sigma(\nu)$ the photoionization cross-sections. In this paper, we use background intensity from Madau \& Haardt 2005 (hereafter MH05) at redshift of $0$, and photoionization cross-sections from \citet{verner96}. Then the ionization rate of $i$th ion for one element can be written as:
  \begin{align}
    R_i \equiv \frac{\partial n_i}{\partial t} = n_\mathrm{e} [n_{i+1} \alpha_{i+1} + n_{i-1} C_{i-1} & - n_i (\alpha
  _i + C_i) ] \notag \\
    & + \left(n_{i-1} \Gamma_{i-1} - n_i \Gamma_i\right),
  \end{align}
  where $n_i$ is the number density of $i$th ion, $n_\mathrm{e}$ the electron number density, $\alpha$ coefficient of radiative and dielectronic recombination, $C$ the coefficient of collisional ionization and auto-ionization, and $\Gamma$ is the coefficient of photoionization. In our simulations, the stiff systems of ordinary differential equations above for each species are solved, adopting the inputs of the current density and temperature, which are updated by the hydrodynamic unit.

\begin{figure}
    \begin{center}
      \includegraphics[width=0.5\textwidth]{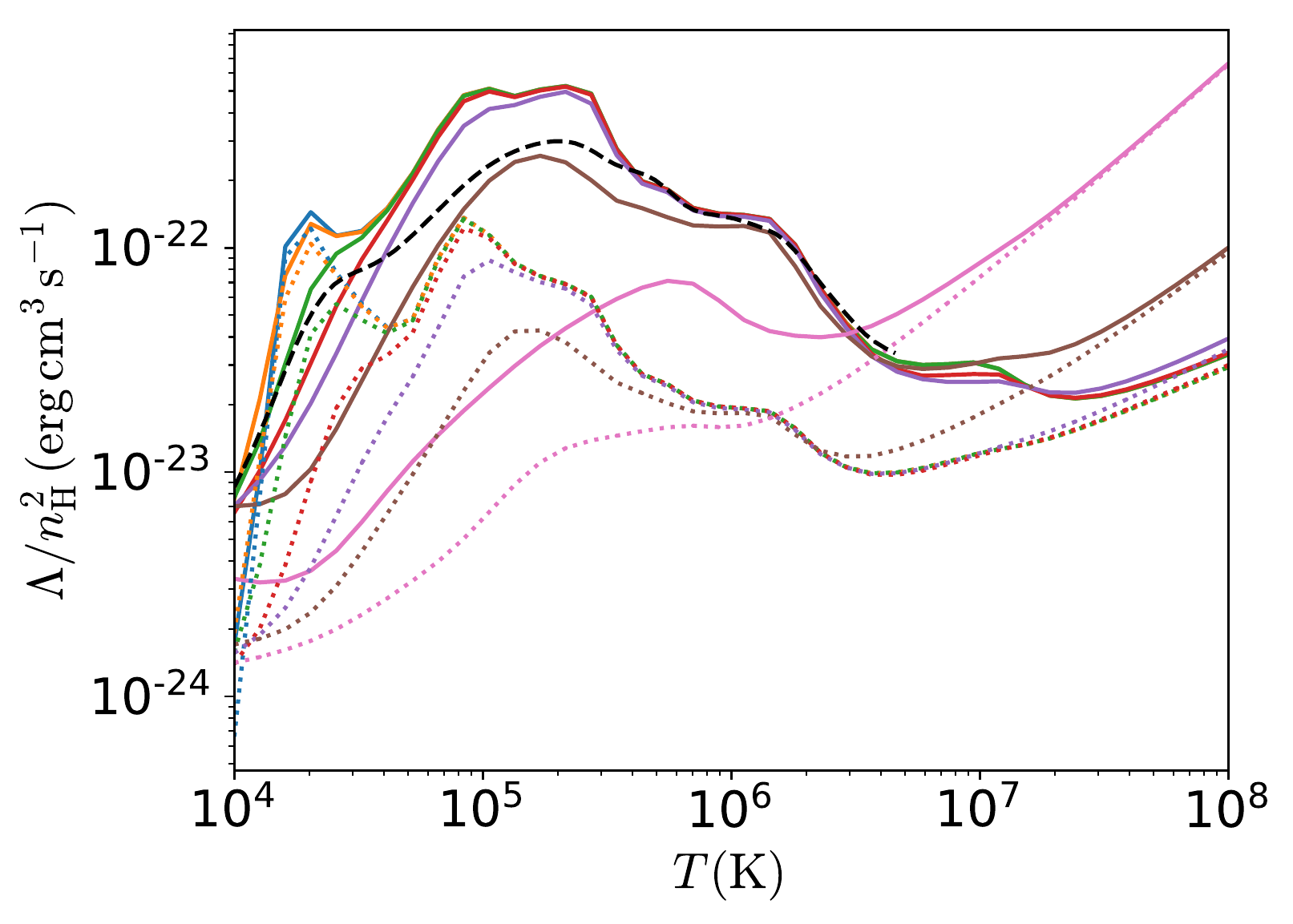}
      \caption{A few selected cooling curves at $Z=Z_\odot$ (solid lines) and $Z=0.1 Z_\odot$ (dotted lines) for number densities $n \, ({\rm cm^{-3}})$ of $1$ (blue), $10^{-1}$ (orange), $10^{-2}$ (green), $10^{-3}$ (red) $10^{-4}$ (purple), $10^{-5}$ (brown) and $10^{-6}$ (pink). The black dashed line is time-dependent cooling curve at $Z=Z_\odot$ without photoionization, which takes NEI effects into account, from \citep{gnat07}.}
      \label{fig:cooling}
    \end{center}
  \end{figure}

  Radiative cooling is also included in our simulations. We use a cooling curve which assumes collisional and photo-ionization ionization equilibrium\footnote{Similarly, \citealt{kwak10}, use a cooling curve which assumes CIE.}, instead of summing up radiative losses from all the time-dependent ion abundances, since the latter requires excessive memory and computation. We run the spectral synthesis code CLOUDY with HM05 background radiation at redshift $0$ to generate the cooling curves with photoionization. Since the cooling curve with photoionization is a function of both temperature and density, we choose $121$ density values logarithmically distributed within the range of $10^{-8}\ \mathrm{cm^{-3}} \leq n \leq 10^1\ \mathrm{cm^{-3}}$, and generate a cooling curve for each density value. Fig. \ref{fig:cooling} shows some cooling curves for selected values of number density and metallicity. These cooling curves are used as input to get the interpolated cooling coefficients at any given density in our simulations. For the conditions considered here, the differences between these curves and net cooling curves (where photoheating is taken into account) is small, and confined to low ($T \sim 2 \times 10^{4}\ \mathrm{K}$) temperatures. 

  To preserve the accuracy of our cooling calculations and avoid the simulation timestep being limited by a very short cooling time, we implement the ``exact'' cooling algorithm described in \citet{townsend09} in the FLASH code. This exploits the fact that exact solutions for radiative cooling exist for a power-law cooling function. We solve the operator-split energy equation with a separation of variables:
  \begin{align}
    t_\mathrm{cool} \frac{dT}{dt} = - T_0 \frac{\Lambda(T)}{\Lambda_0},
  \end{align}
  where $t_\mathrm{cool}$ is cooling time, $\Lambda(T)$ the cooling coefficient expressed as a piecewise power law function of $T$, $T_0$ and $\Lambda_0$ are arbitrary reference temperature and cooling coefficient. The piecewise power law functions are attained by dividing each cooling curves at a certain density into $40$ temperature bins and performing a linear fit in each bin.

  Since simulations in this paper span over a wide range of parameters and their combinations, it is redundant to explicitly list every individual simulations with the parameters used. Instead, here we attach a number of name elements to corresponding parameter values which are described in Tab. \ref{tb:parameter}, and use a combination of these name elements to refer to one simulation where a certain combination of parameters are used. For instance, the name ``{\tt Z0.1\_nei\_photo\_Bx-lgr}'' refers to the NEI simulation with $Z = 0.1 Z_\odot$, strong magnetic fields of $B_x = 0.35 \ \mathrm{\mu G}$, photoionization included and other parameters the same as fiducial ones. Note that to isolate the role of cooling, in some simulations we elect to change the normalization of the cooling curve by a constant prefactor. 

\section{Results}
\label{sect:results}

\subsection{Hydrodynamic simulations}

  We first study purely hydrodynamic mixing layers in collisional ionization equilibrium. Surprisingly, although the analytic estimate in Eq. \eqref{eq:TML_column} has been compared against observations to infer physical conditions (e.g., \citealt{slavin93,bordoloi16}), it has never been compared directly against numerical simulations, where all physical quantities are known. In fact, we shall see that naive application of Eq. \eqref{eq:TML_column}, assuming that the turbulent velocity is of order the shear velocity $v_\mathrm{t}\sim v$, is inaccurate and gives incorrect scalings with physical quantities.

\begin{figure*}
    \begin{subfigure}[b]{\textwidth}
      \includegraphics[width=\textwidth]{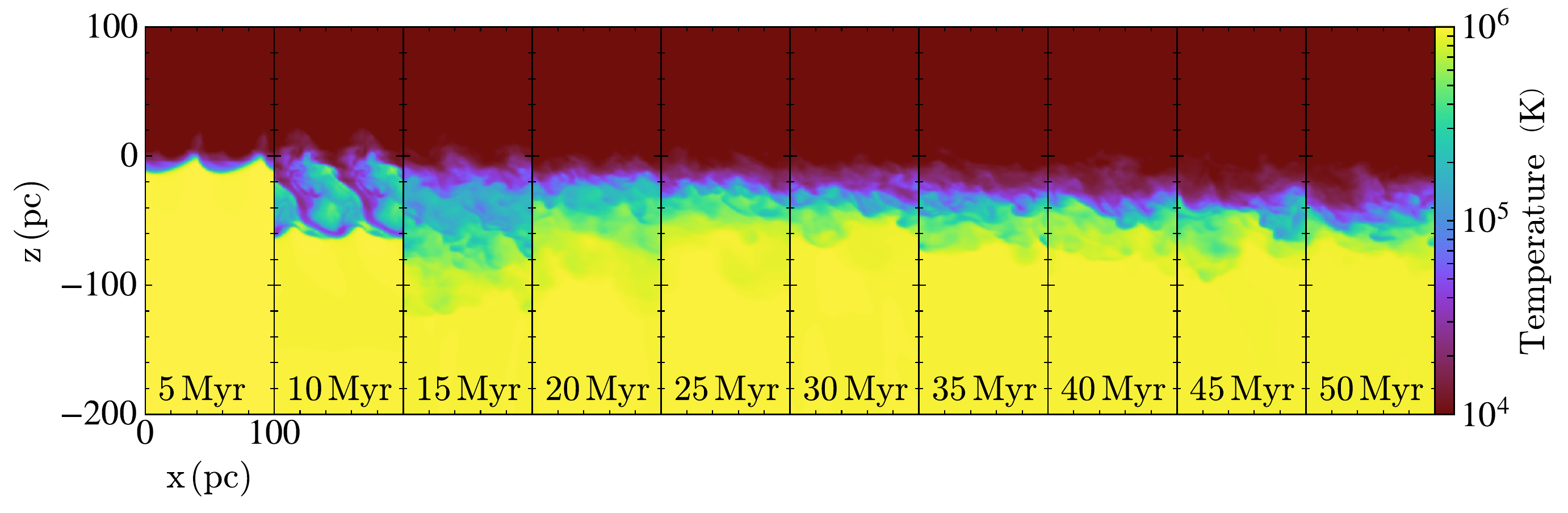}
    \end{subfigure}
    \begin{subfigure}[b]{\textwidth}
      \includegraphics[width=\textwidth]{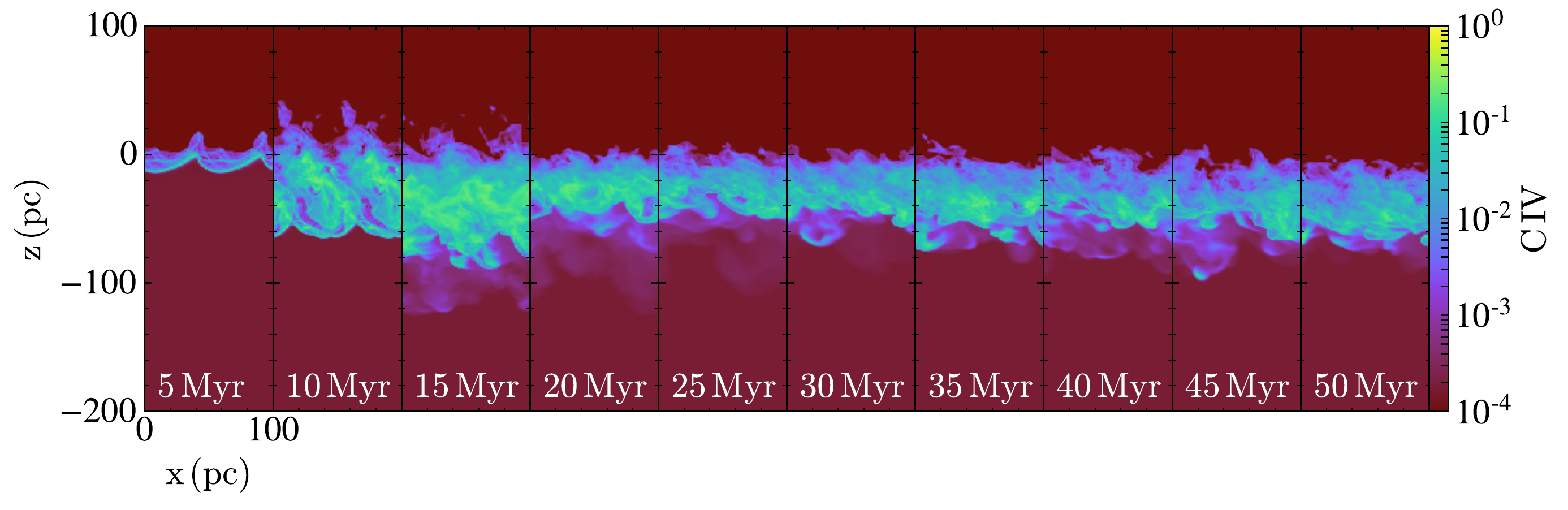}
    \end{subfigure}
    \begin{subfigure}[b]{\textwidth}
      \includegraphics[width=\textwidth]{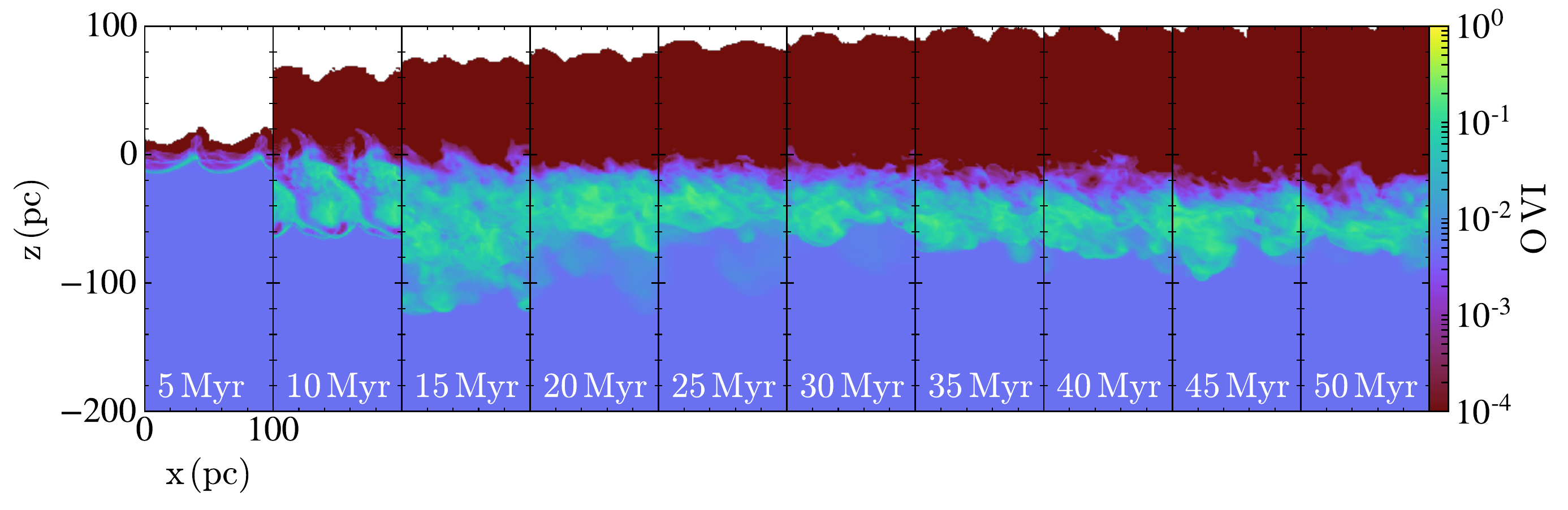}
    \end{subfigure}
    \begin{subfigure}[h]{\textwidth}
      \includegraphics[width=\textwidth]{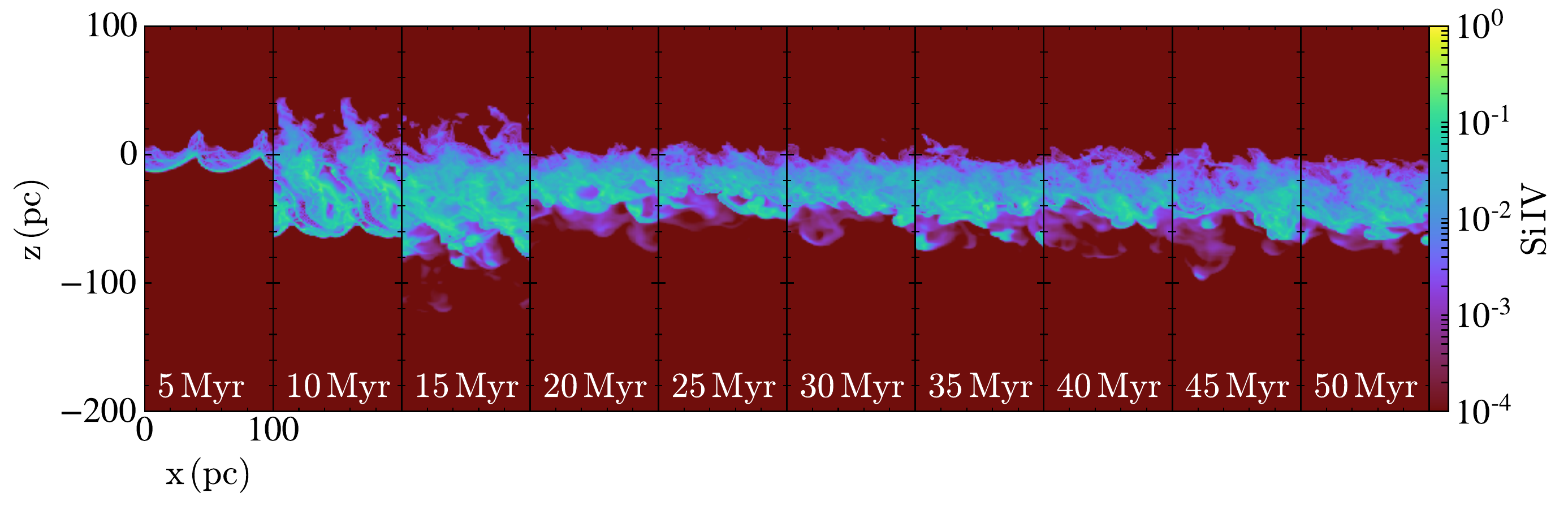}
    \end{subfigure}
    \caption{The timeseries of density-weighted projections of temperature and \CIV, \OVI and \SiIV fractions, from CIE hydrodynamic simulations {\tt Z1}.}
    \label{fig:proj_CIE}
  \end{figure*}

\begin{figure*}
    \begin{center}
      \includegraphics[width=\textwidth]{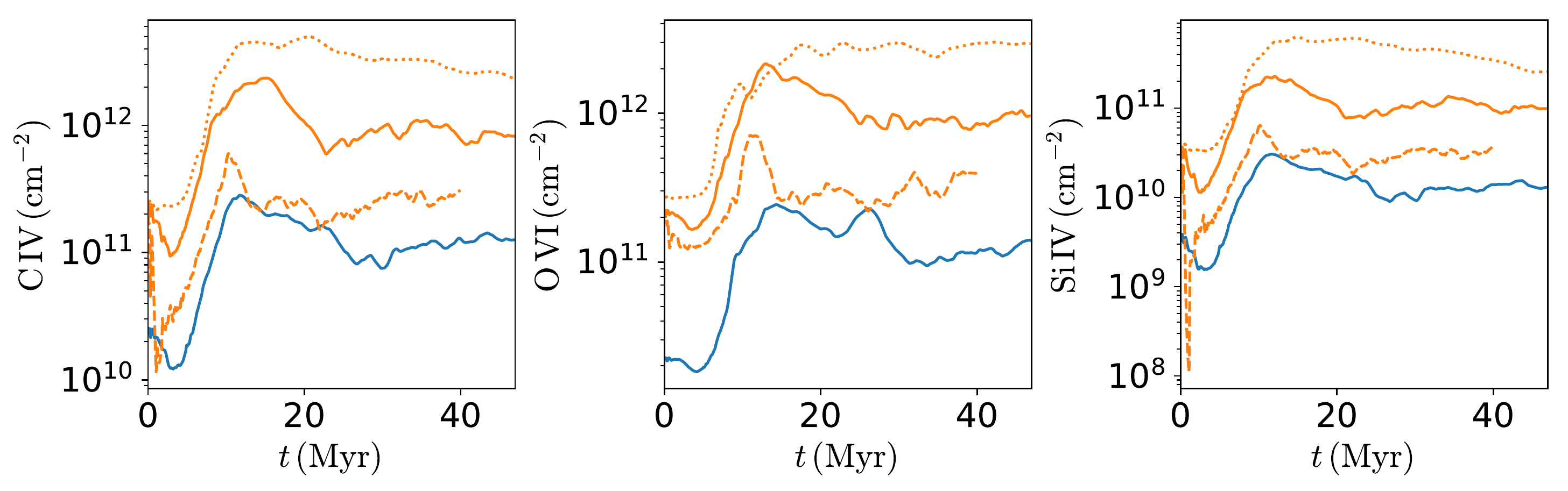}
      \caption{Effect of metallicity and cooling on ion column densities: solid blue --- {\tt Z0.1}, solid orange --- {\tt Z1}, dotted orange --- {\tt Z1\_cool0.1}, dashed orange --- {\tt Z1\_cool10}. Lower metallicities depress column densities in CIE hydrodynamic simulations, in contrast with analytic predictions (\ref{eq:TML_column}). Column densities do decrease as $t_{\rm cool}$ falls, but with a weaker dependence than suggested by equation (\ref{eq:TML_column}).}
      \label{fig:eqi_Z_cool}
    \end{center}
  \end{figure*}

\begin{figure*}
    \begin{center}
      \includegraphics[width=\textwidth]{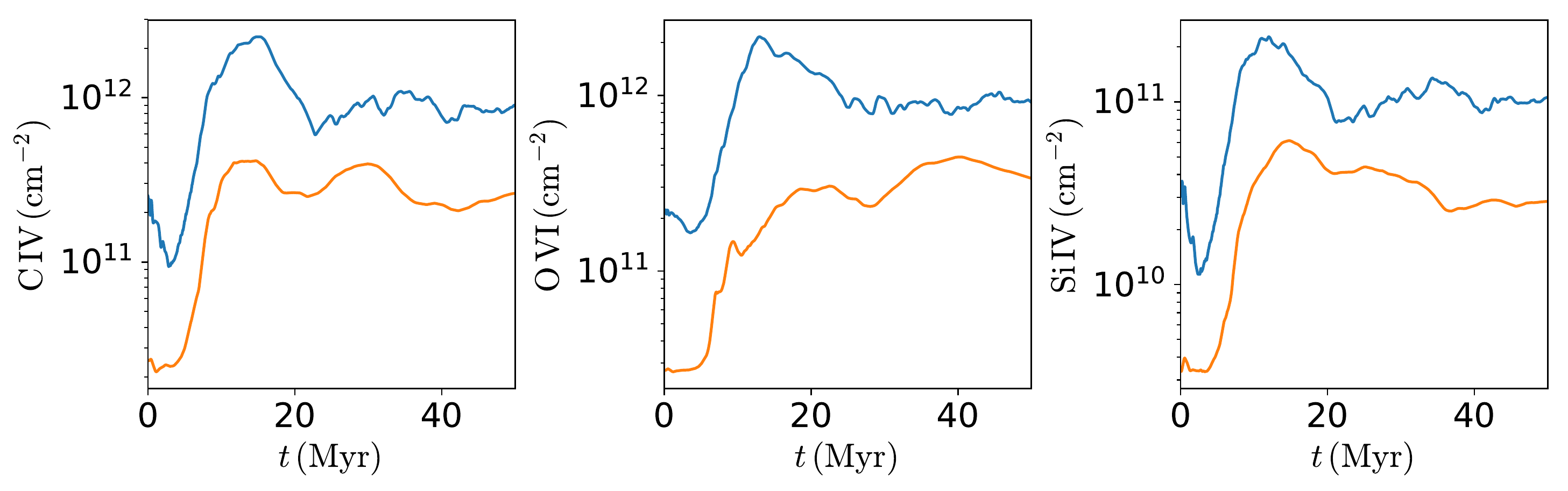}
      \caption{Effect of ambient pressure on ion column densities: blue --- {\tt Z1}, orange --- {\tt Z1\_pres-sml}. Column densities decrease as pressure (and hence density) decreases, contrary to the expectation from Eq. \eqref{eq:TML_column} that column densities are independent of density.}
      \label{fig:eqi_lowpres}
    \end{center}
  \end{figure*}

\begin{figure*}
    \begin{center}
      \includegraphics[width=\textwidth]{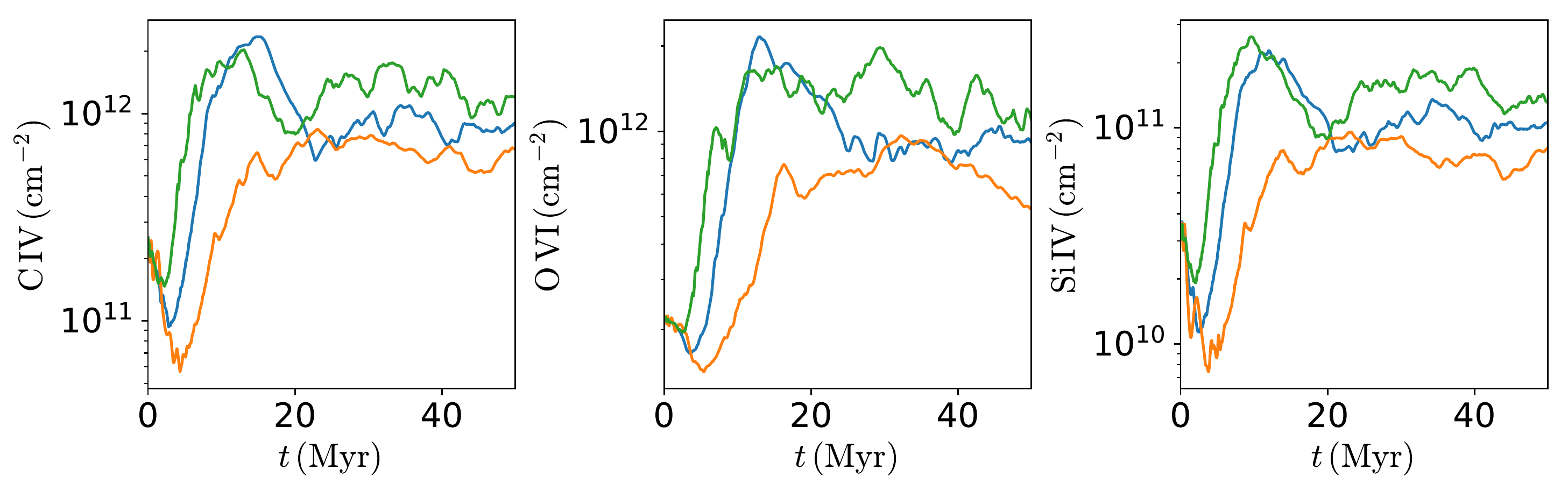}
      \caption{Effect of initial shear velocity on ion column densities: blue --- {\tt Z1}, orange --- {\tt Z1\_vel-sml}, green --- {\tt Z1\_vel-lgr}. Column densities only have a very weak sublinear dependence on shear.}
      \label{fig:eqi_shear}
    \end{center}
  \end{figure*}

\begin{figure*}
    \begin{center}
      \includegraphics[width=\textwidth]{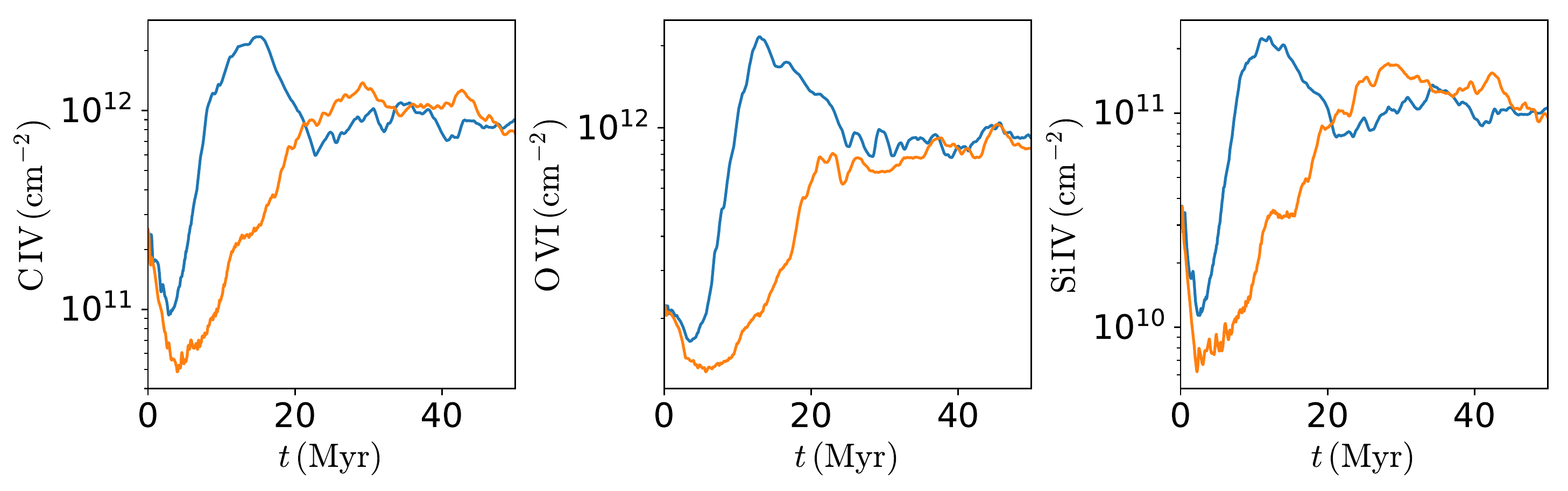}
      \caption{Effect of density contrast on ion column densities: blue --- {\tt Z1}, orange --- {\tt Z1\_contr-lgr}. Steady-state column densities appear independent of density contrast, even if the transient behavior differs due to different mixing timescales.}
      \label{fig:eqi_overdens}
    \end{center}
  \end{figure*}

  Fig. \ref{fig:proj_CIE} shows the density-weighted projections of temperature and \CIV, \OVI, \SiIV fractions in the run {\tt Z1}, whose ion column densities are shown by the solid orange curve In Fig. \ref{fig:eqi_Z_cool}. The column densities stabilize at $\sim 10^{12}\ \mathrm{cm^{-2}}$ for \CIV and \OVI, and $\sim 10^{11}\ \mathrm{cm^{-2}}$ for \SiIV. To within an order of magnitude, the \OVI column density from our simulations is roughly consistent with Eq. \eqref{eq:TML_column}, assuming the fudge factor $\eta \sim 0.1$. Eq. \eqref{eq:TML_column} also makes several predictions for functional dependences: since $t_\mathrm{cool}\propto 1/Zn$ in the crucial temperature range, in CIE the column density is roughly independent of the underlying density (or pressure, assuming isobaric conditions) and metallicity, while the column density is proportional to both the cooling time $t_\mathrm{cool}$ and (assuming $v_\mathrm{t} \sim v$) the shear velocity $v$. In Fig. \ref{fig:eqi_Z_cool}, \ref{fig:eqi_lowpres} and \ref{fig:eqi_shear}, we see that many of these expectations are not borne out. Fig. \ref{fig:eqi_Z_cool} shows that contrary to expectations, the column density is directly proportional to metallicity $N\propto Z$: between the $Z=0.1 Z_\odot$ (solid blue) and $Z = Z_\odot$ (solid orange), the column density increases by an order of magnitude. Fig. \ref{fig:eqi_Z_cool} also shows the effect of changing the cooling time: the dotted and dashed orange lines shows results when the normalization of the cooling curve $\Lambda(T)$ is artificially reduced (increased) by a factor of $0.1$ ($10$). The ionic column densities do decrease, but only by a factor of $3$ instead of a factor of $10$. The orange line in Fig. \ref{fig:eqi_lowpres} shows the column densities in the hydrodynamic case when the pressure (and hence density) has been reduced by a factor of $10$, from $P/k_\mathrm{B}=160 \ \mathrm{cm^{-3} K}$ to $P/k_\mathrm{B}=16 \ \mathrm{cm^{-3} K}$. Even in the absence of photoionization, which introduces density dependence via the ionization parameter, CIE column densities in the low density/pressure case are a factor of a few smaller than in the fiducial case. In Fig. \ref{fig:eqi_shear}, we explore the effect of different shear velocities. The column density does decrease with shear, though much more weakly compared to the expected linear scaling.\footnote{Note that transonic flows are most relevant to clouds in galaxy halos, where orbital velocities are comparable to the hot gas sound speed. We shall explore the effects of highly supersonic shear, which is most relevant to jets, elsewhere.} Finally, in Fig. \ref{fig:eqi_overdens}, we explore the role of density contrast $\delta$ comparing the $\delta=100$ and $\delta=1000$ cases. We find that the steady-state column densities are independent of $\delta$, although transient behavior differs due to differing mixing timescales. 

What could have gone wrong? Eq. \eqref{eq:TML_column} is straightforward and well-known to be accurate in radiative shocks (where $v$ is the post-shock velocity). There are two possibilities. One, our estimate of $t_\mathrm{cool} (n,T)$ is incorrect, either due to incorrect assumptions about the characteristic temperature $T_{\rm mix} \sim\sqrt{T_\mathrm{hot} T_\mathrm{cold}}$ or density ($n \sim P/k_\mathrm{B} T_{\rm mix}$, assuming isobaricity) at which it should be evaluated.\footnote{For high ions such as \OVI, the ions peak at a characteristic temperature of $T\sim10^{5.5}\ \mathrm{K}$ with a fairly narrow speed, which removes ambiguity about the appropriate temperature.} Two, our estimate of $v\sim v_\mathrm{turb}\sim v_\mathrm{shear}$ is incorrect. The usual argument \citep{begelman90} is that $v_\mathrm{turb}$ is dominated by the velocity of the largest eddy, which is of order the shear velocity. Indeed, this is true for the adiabatic Kelvin-Helmholtz instability. However, examination of our simulation outputs show that our radiative turbulent mixing layers have turbulent velocities much less than $v_\mathrm{shear}$, by $1$ --- $2$ orders of magnitude. This is the dominant reason for the discrepancy with Eq. \eqref{eq:TML_column}.

There are two obvious reasons why turbulent velocities should be less than the shear velocity. Firstly, as the hot gas mixes with and entrains cold gas, by momentum conservation, its velocity falls. Secondly, the shear velocity is highly supersonic in the cool mixing region (which has a considerably lower sound speed); radiative shocks thus quickly dissipate motions. Indeed, our simulations indicate frequent shocks in the mixing layer. However, the transverse flow velocities seen in the simulations are very small ($\sim 5\ \mathrm{km/s}$ for the $Z=Z_\odot$ case), and thus subsonic rather than transonic --- even in the mixing layer.

In order to understand this, we first consider the leading order terms in the fluid equations, then apply them to our simulations. 

    \begin{figure}
    \begin{center}
      \includegraphics[width=0.5\textwidth]{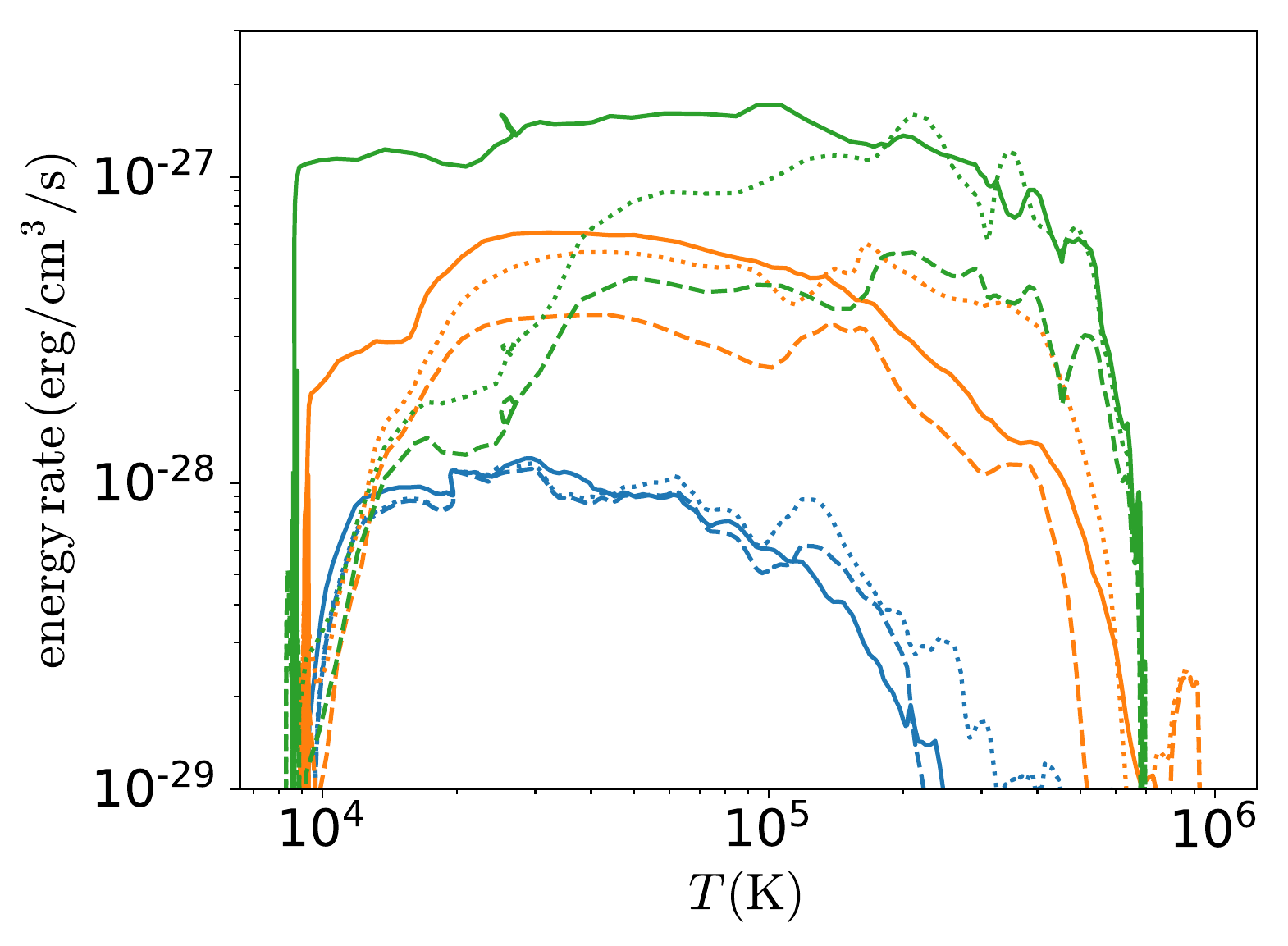}
      \caption{Profiles of emissivity $n_\mathrm{e} n_\mathrm{i} \Lambda$ (solid lines), the divergence of energy flux $-\nabla_y 5/2 P v_z$ (dashed lines), and the sum of turbulent dissipation $-\left\langle\rho v_x^\prime v_z^\prime \right\rangle \partial_z\tilde{v}_x$ and energy flux divergence (dotted lines) from CIE hydrodynamic simulations with different cooling rates at $t=50 \ \mathrm{Myr}$: blue --- {\tt Z1\_cool0.1}, orange --- {\tt Z1}, green --- {\tt Z1\_cool10}.}
      \label{fig:energy_flux}
    \end{center}
  \end{figure}
  
\subsubsection{Order of Magnitude Analysis of Fluid Equations}
\label{sec:OOM}

Here, we review classical order of magnitude analysis of mixing layers \citep{tennekes1972first,white2015turbulent}, which is useful for understanding their structure. The reader who is not interested in these details should skip ahead to \S\ref{sect:hydro_analysis}.

We consider the time stationary equations of mass, momentum (both streamwise and transverse) and energy conservation respectively, where we decompose fluid variables into a mean and fluctuating component respectively. If one adopts a convention whereby extensive variables such as density are time or ensemble averaged ($\rho=\bar{\rho} + \rho^\prime; \left\langle\rho^\prime\right\rangle=0$) but intensive variables such as velocity are mass averaged ($v_i = \tilde{v}_i + v_i^\prime; \left\langle \rho v_i^\prime\right\rangle=0$), then the time-independent hydrodynamic conservation equations are \citep{favre1969problems,kuncic2004dynamics}:

\begin{gather}
  \frac{\partial}{\partial x_j} \left(\bar{\rho}\tilde{v}_j\right) = 0 \\
  \frac{\partial}{\partial x_j} \left(\bar{\rho}\tilde{v}_i \tilde{v}_j\right) = -\frac{\partial \bar{P}}{\partial x_i} + \frac{\partial}{\partial x_j}\left(t_{ij}^R\right) \\
  \frac{\partial}{\partial x_i} \left(\bar{u}\tilde{v}_i +\left\langle u v_i^\prime\right\rangle\right) = - \bar{P} \tilde{v}_{i,i} - \left\langle P v_{i,i}^\prime\right\rangle - n^2 \Lambda(T) - t_{ij}^R \tilde{S}_{ij}
\end{gather}
where the Reynolds stress is:
\begin{align}
  t_{ij}^R = - \left\langle\rho v_i^\prime v_j^\prime\right\rangle,
\end{align}
the mean fluid shear is:
\begin{align}
  \tilde{S}_{ij} = \frac{1}{2} \left(\tilde{v}_{i,j} + \tilde{v}_{j,i} - \frac{2}{3} \delta_{ij} \tilde{v}_{k,k}\right).
\end{align}
As in \citet{kuncic2004dynamics}, we use commas to denote partial derivatives, and adopt the Einstein summation correction. The momentum equation denotes that momentum transport is accomplished by pressure gradients and the Reynold stress. The energy equation denotes that the change in the gas internal energy (left hand side) is due to adiabatic compression/expansion in the mean and fluctuating flow, radiative cooling, and turbulent dissipation (respective terms on the right hand side). For the last term, we have made the usual ``on the spot'' assumption, equating the production rate of turbulence (given by the work done by shear in the mean flow on Reynolds stresses) to the dissipation rate, neglecting any non-local or transport effects.

It is useful to eliminate unimportant terms in these equations via an order-of-magnitude analysis in the two physically distinct dimensions $x$ and $z$, averaging over the $y$ direction \citep{tennekes1972first,white2015turbulent}. Following \citet{white2015turbulent}, we denote characteristic length scales in the $x$ and $z$ directions as $L$ and $h$, where $h\ll L$ by assumption of a thin mixing layer. The former is of order the thickness of the mixing layer, while the latter can be thought of as the length scale on which the flow accelerates or decelerates. The equation of continuity
\begin{align}
  \frac{\partial}{\partial x}\left(\bar{\rho}\tilde{v}_x\right) + \frac{\partial}{\partial z}\left(\bar{\rho}\tilde{v}_z\right) = 0
\end{align}
thus yields
\begin{align}
  \tilde{v}_z \sim \frac{h}{L} \tilde{v}_x.
\end{align}
If we use this relation in the momentum equation, and the fact that $h\ll L$, the leading order terms in the streamwise momentum equation are:
\begin{align}
  \frac{\partial}{\partial x}\left(\bar{\rho}\tilde{v}_x^2\right) + \frac{\partial}{\partial z} \left(\bar{\rho}\tilde{v}_x \tilde{v}_z\right) = - \frac{\partial}{\partial x} \left\langle\rho v^\prime_x v^\prime_z\right\rangle.
  \label{eq:steam_mom}
\end{align}
Note that we have retained the term involving the Reynolds stress, even though we have not yet established its scaling. We cannot eliminate it, otherwise there is no momentum transport! Indeed, if we assume roughly isotropic turbulence, $v_x^\prime \sim v_z^\prime \sim v_\mathrm{t}$, Eq. \eqref{eq:steam_mom} thus tells us from balancing the left and right hand sides that
\begin{align}
  v_\mathrm{t}^2 \sim \frac{h}{L}v^2.
  \label{eq:vt_scaling}
\end{align}
The leading terms in the transverse momentum equation give:
\begin{align}
  \frac{\partial \bar{P}}{\partial z} + \frac{\partial}{\partial z}\left\langle \rho v^{\prime 2}\right\rangle = 0,
\end{align}
or
\begin{align}
  \bar{P} + \left\langle \rho v^{\prime 2}_y\right\rangle = \text{const}, 
  \label{eq:pressure_const}
\end{align}
which states that the sum of thermal and turbulent pressure is approximately constant.

Finally, the leading order terms in energy equation give:
\begin{align}
  \frac{\partial}{\partial x}\left(\bar{u}\tilde{v}_x\right) + \frac{\partial}{\partial z}\left(\bar{u}\tilde{v}_z\right) = - \bar{P} \left(\frac{\partial v_x}{\partial x} + \frac{\partial v_z}{\partial z}\right)&  -  n^2 \Lambda(T) \notag \\ 
    &- \left\langle\rho v_x^\prime v_z^\prime \right\rangle \frac{\partial\tilde{v}_x}{\partial z}.
  \label{eq:energy}
\end{align}

The next step is to integrate out the slow variation of the mixing layer in the $x$-direction (which in any case is not present in our simulations, which have periodic boundary conditions in the $x$-direction), to determine the average vertical structure of the mixing layer in the $z$ direction. However, there are two stumbling blocks. One, the fluid equations (mass conservation, two momentum equations, and the energy equation) are 4 equations in 5 variables $\tilde{n}, \tilde{T}, \tilde{v}_x, \tilde{v}_y, \tilde{v}_t$. Secondly, we have a missing crucial boundary condition: the effective entrainment velocity $\tilde{v}_z$ as $z \to \pm \infty$, which determines the mass and enthalpy flux into the mixing layer. To date, analytic models of the mixing layers around jets either use experimentally calibrated entrainment efficiencies\footnote{Note, however, that the experiments involve adiabatic mixing layers; using these entrainment efficiencies for radiative mixing layers is questionable. Indeed, our simulations show that radiative cooling, rather than the Kelvin Helmholtz instability, is paramount in setting the entrainment rate.} \citep{canto1991mixing,raga1995mixing}, or use observations of the growth of the mixing layer with distance (or effectively the opening angle of the mixing layer, e.g. \citet{white2015turbulent}, which corresponds to the parameter $\alpha$ below) to determine the entrainment efficiency. In the following section, we see how our simulations can provide clues as to closing this system of equations.

\subsubsection{Analysis of Hydrodynamic Simulations}
\label{sect:hydro_analysis}

We can make progress as follows:

{\bf Energetics} As evident in the energy equation (Eq. \eqref{eq:energy}), there are 3 possible ways to offset radiative losses in the mixing layer: advection of thermal energy (first term on LHS), adiabatic compression, or dissipation of turbulence (first and last term on RHS respectively). For instance, \citet{white2015turbulent} assume turbulent dissipation balances cooling. Advection of thermal energy taps the free energy of the high entropy hot gas reservoir, while turbulent dissipation taps the free energy of the shear flow. In Fig. \ref{fig:energy_flux}, we compare the leading terms in the energy equation (radiative cooling, divergence of enthalpy flux, and dissipation of turbulence), for solar metallicity, and where the cooling function is rescaled by a factor of 0.1 and 10. These terms are averaged over slices of the mixing layer (i.e., at fixed z), and compared against the average temperature in that layer. We observe the following: 
\begin{itemize}

\item{The emissivity $n_{\rm e} n_{\rm i} \Lambda(T)$ (solid lines) is not what one would expect for isobarically cooling gas as a function of temperature $T$. Instead, it has a much lower normalization and a much flatter temperature dependence (we expect $\epsilon \propto \Lambda(T)/T^{2}$ to vary by $2$ -- $3$ orders of magnitude over the temperature range of interest). Also, the emissivity increases more weakly with the cooling function $\Lambda(T)$ than expected (an order of magnitude increase in $\Lambda(T)$ only yields a factor $\sim 3$ increase in emissivity).}

\item{The sum of heating terms is sufficient to balance radiative cooling for $T > T_{\rm crit}$ (compare solid and dotted lines), where $T_{\rm crit}$ is some critical temperature, but energy balance fails at lower temperatures. Moreover, $T_{\rm crit}$ increases as the normalization of $\Lambda(T)$ increases. Finally, turbulent dissipation becomes increasingly important as the normalization of $\Lambda(T)$ increases.}

\end{itemize} 

Closer inspection of the simulations allows us to qualitatively understand these trends. The gas does not cool monolithically but remains highly multi-phase through the mixing layer; what changes through the mixing layer is the relative amounts of cold and hot gas (which result in a change in mean temperature). Most of the volume is filled with such gas, which has low emissivity; this accounts for the reduced overall emissivity compared to single-phase expectations. What dominates the emission is the ``skin'' of mixed gas at intermediate temperatures surrounding finger-like structures in the mixing layer. This is most clearly evident in slice (rather than projection) plots. Since these structures have sizes of order the mixing layer depth, the filling fraction and emissivity of such gas varies only weakly with mixing layer depth. Otherwise, strong variations in emissivity will be counteracted by enthalpy flux. The sub-linear increase of emissivity with $\Lambda(T)$ is due to the shorter lifetime of mixed gas at higher $\Lambda(T)$, resulting in smaller amounts of emitting gas. 

The power source for the mixing layer is the enthalpy flux from the hot gas, in addition to dissipation of shear motions. Near the hot end, they can keep the mixing layer in energy balance, but get weaker as one moves towards the cold gas. At some point towards the cold end of the mixing layer, they are unable to overcome the stronger radiative cooling and wind of low entropy gas entrained from the cold reservoir. At this point, corresponding to a mean temperature $T_{\rm crit}$, the mixing layer falls out of energy balance and cools passively and monolithically. This happens at higher $T_{\rm crit}$ for stronger cooling (higher $\Lambda(T)$). From Eq. \eqref{eq:pressure_const}, with stronger cooling we also expect increased turbulence, due to the faster drop in thermal pressure. If the cooling time is shorter than the sound crossing time, the ``piston'' of over-pressured hot gas pushing into the cooling region generates turbulence. Also, the turbulence can be seeded by interaction of the mean flow with density inhomogeneities created by cooling. In our simulations, as cooling increases, we see more weak shocks, stronger dissipation, and increased turbulence. 

{\bf Entrainment rates} The cooling flow causes the cooling column to shrink. The shrinkage cannot be communicated at a velocity higher than the sound speed of the freely cooling gas, $c_{\rm s}(T_{\rm crit})$. This shrinkage sets the entrainment rate of the hot gas $\tilde{v}_{z}$. We can quantify these rates from our simulations. When energy balance applies, we expect: 
\begin{equation}
\tilde{v}_{z}^{\rm hot} \left[ \frac{5}{2} P \left(1 + \mathcal{M}^{2}\right) \right] \approx \Sigma_{\rm L} =  \int_0^h dz n^2(z) \Lambda(T)
\label{eq:energy_balance}
\end{equation}
where $\mathcal{M}$ is the Mach number of the flow, and $\tilde{v}_{z}^{\rm hot}$ is measured in a frame where the mixing layer is stationary\footnote{In our boxes, particularly with strong cooling, the mixing layer travels downwards into the hot phase, and this velocity must be added to the hot gas inflow rate at the boundary to obtain $\tilde{v}_{z}^{\rm hot}$.}. The surface brightness is balanced by the advected enthalpy and bulk kinetic energy, which is assumed to be dissipated. In our simulations, keeping in mind that we have only explored a limited portion of parameter space with a small number of simulations, a fairly good fit at solar metallicity is given by: 
\begin{equation}
\Sigma_{\rm L} \approx 6 \times 10^{-8} {\rm erg \, s^{-1} \, cm^{-2}} v_{100}^{1/4} \left[\Lambda_{-21.4}^{\rm T_{\rm mix}}\right]^{1/2} P_{-13.7}^{3/2}
\end{equation}
independent of overdensity $\delta$, where the cooling function is evaluated at $T_{\rm mix} \sim 10^{5}$K, and $v_{100} \equiv (v_{\rm x}/100 \, {\rm km \, s^{-1}})$, $\Lambda_{-21.4}^{\rm T_{\rm mix}}\equiv \left(\Lambda(T_{\rm mix})/4 \times 10^{-22} \, {\rm erg \, s^{-1} \, cm^{3}}\right)$, and $P_{-13.7} \equiv \left(P/2.2 \times 10^{-14} \, {\rm erg \, cm^{-3}}\right)${, which can be translated to $P/k_\mathrm{B} = 145\ \mathrm{cm^{-3}K}$}. \footnote{For references, the mean thermal pressure in galactic disk is $P/k_\mathrm{B} = 3800\ \mathrm{cm^{-3}K}$ \citep{jenkins2011distribution}, and in galactic halo absorbers is $P/k_\mathrm{B} = 1\ \text{--}\ 30\ \mathrm{cm^{-3}K}$ \citep{keeney2017characterizing}.} Variations in cooling due to change in metallicity, photoionization, etc, can be approximately incorporated by the scaling in the cooling function normalization. From Eq. \eqref{eq:energy_balance}, this implies an inflow velocity of: 
\begin{equation}
\tilde{v}_{z}^{\rm hot} \sim 8 \, {\rm km \, s^{-1}} v_{100}^{1/4} \left[\Lambda_{-21.4}^{
\rm T_{\rm mix}}\right]^{1/2} P_{-13.7}^{1/2} \left(\frac{1+ \mathcal{M}^{2}}{1.5}\right)^{-1}
\label{eq:vy}
\end{equation}
which indeed is a good match to what we measure in the simulations, implying that Eq. \eqref{eq:energy_balance} holds.\footnote{As a side note, the inflow velocity in the frame of the box boundary $v_{z, \mathrm{box}}^\mathrm{hot}$ follows an even weaker dependence on $\Lambda(T)$ with the power-law index roughly between $1/4$ and $1/3$, which reflects the mass accretion rate onto the cold phase.} To within a factor of $2$, $\tilde{v}_{z}^{\rm hot} \sim c_{\rm s} (T_{\rm crit})$, where $T_{\rm crit} \sim 1.5 \times 10^{4} \, {\rm K}$ in our fiducial case. The fact that $\tilde{v}_{z}^{\rm hot}$ is independent of $\delta$ and only weakly dependent on $v_{\rm x}$ indicates that entrainment of hot gas into the mixing layer is set by radiative cooling rather than the Kelvin-Helmholtz instability (where growth rate is set by these two parameters). The scaling with $\Lambda$, $P$ indicates that $v_z^\mathrm{hot} \propto t_\mathrm{cool}^{-1/2}$ (since $t_\mathrm{cool} \simeq (k_{\rm B} T_\mathrm{mix})^{2} / P \Lambda(T_\mathrm{mix})$ at constant pressure). It is reasonable that $v_z$ increases with a  shorter cooling time, since a higher enthalpy flux is required to offset cooling, but we can only obtain the exponent numerically. Equation can be used to determine the crucial dimensionless parameter 
\begin{align}
    \alpha \sim \frac{\tilde{v}_z}{\tilde{v}_x} \sim \frac{h}{L} \sim \frac{c_{\rm s}(T_{\rm crit})}{\tilde{v}_{\rm x}}.
\label{eq:alpha_define}
  \end{align}
which determines the overall width and column density of the mixing layer. It is set by the non-linear saturation of the radiative Kelvin-Helmholtz instability, and requires numerical simulation to be accurately calculated, but the last equality above enables us to have an order of magnitude understanding of the value $\alpha \sim 0.05$ seen in the simulations. 

The parameter $\alpha$ sets the hot gas entrainment rate. The cold gas entrainment rate is set by the Kelvin-Helmholtz instability \citep{begelman90}:   \begin{align}
    \frac{\dot{m}_\mathrm{hot}}{\dot{m}_\mathrm{cold}} \sim \sqrt{\frac{\rho_\mathrm{hot}}{\rho_\mathrm{cold}}},
  \end{align}
 or $v_z^\mathrm{hot}/v_z^\mathrm{cold} \sim \sqrt{\rho_\mathrm{cold}/\rho_\mathrm{hot}}$. We have verified that this relation is obeyed in our simulation. Note thus that the enthalpy flux into the mixing layer is dominated by hot gas, while the mass flux is dominated by cold gas. Entrainment of hot gas into the mixing layer is driven by radiative cooling: hot gas which enters the mixing layer cools, losing pressure and shrinking in volume. More hot gas thus flows in to take its place. The characteristic inflow rate $v_z^\mathrm{hot} \sim h/t_\mathrm{cool}$ is thus set by cooling. On the other hand, cold gas is drawn into the mixing layer via the Kelvin-Helmholtz instability, with an inflow rate $v_z^\mathrm{cold} \sim \sqrt{\rho_\mathrm{hot}/\rho_\mathrm{cold}} v_z^\mathrm{hot}$ set by the Kelvin-Helmholtz time (which also depends on radiative cooling via $v_z^\mathrm{hot}$).

{\bf Structure of the mixing layer} The characteristic width of the mixing layer is given by: 
\begin{equation}
h \sim \frac{\Sigma_{\rm L}} {\left\langle n_\mathrm{e} n_\mathrm{i} \Lambda (T_{\rm mix}) \right\rangle} \sim \tilde{v}_{\rm z}^{\rm hot} \langle t_{\rm cool} \rangle \sim 50 \, {\rm pc} \left[\Lambda_{-21.4}^{\rm T_{\rm mix}}\right]^{-1/2} P_{-13.7}^{-1/2}
\label{eq:width}
\end{equation} 
Note that $h \propto t_{\rm cool}^{1/2}$, and thus grows more slowly with cooling time than expected, due to the fact that $\tilde{v}_{\rm z}^{\rm hot} \propto t_{\rm cool}^{-1/2}$. Also note that since the emissivity $\langle n_\mathrm{e} n_\mathrm{i} \Lambda (T_{\rm mix}) \rangle$ is smaller than in a single phase model due to the low filling fraction of mixed gas, the width of the mixing layer is substantially larger than $ \tilde{v}_{\rm z}^{\rm hot} t_{\rm cool}(T_{\rm mix})$. 

\begin{figure}
    \begin{center}
      \includegraphics[width=0.5\textwidth]{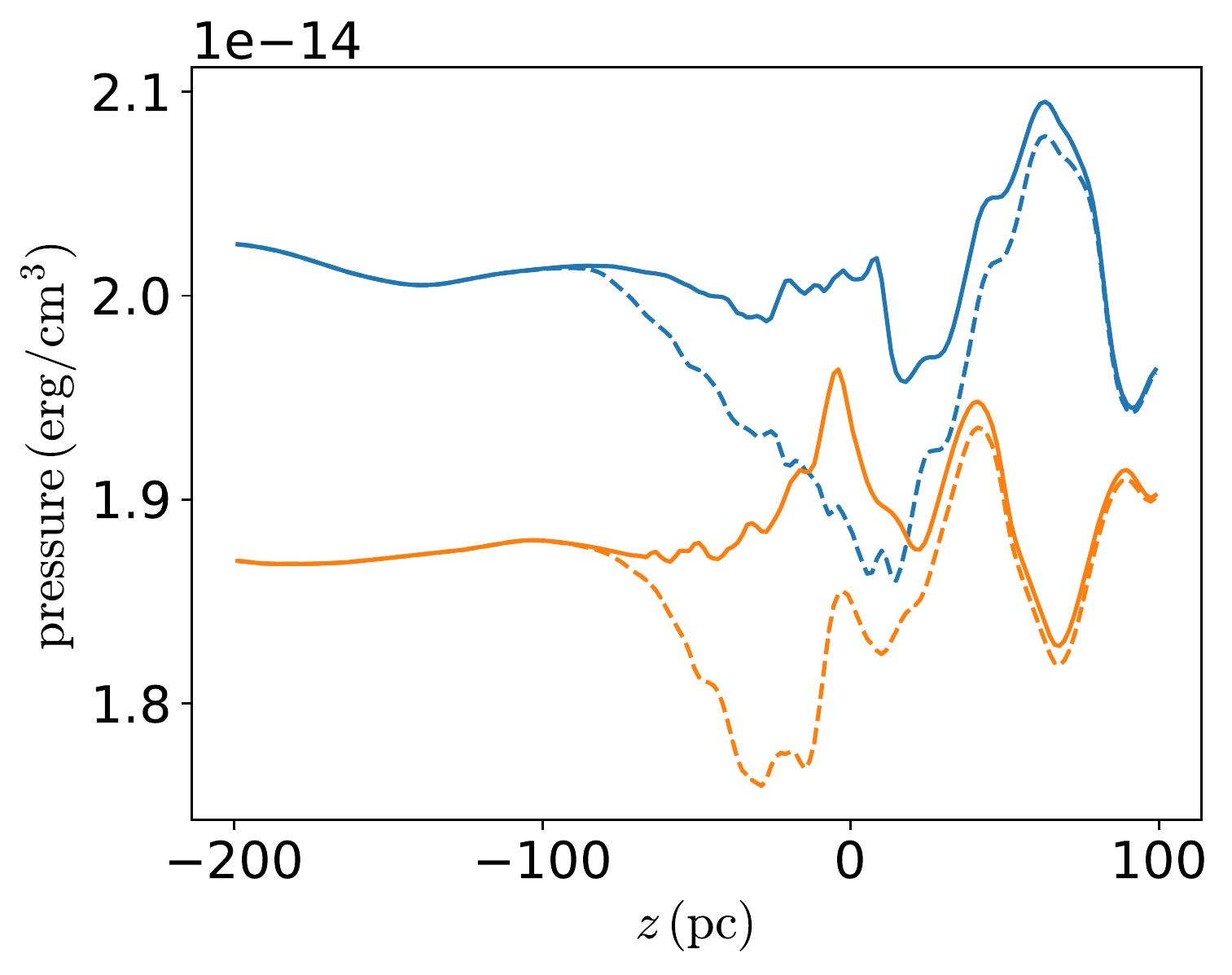}
      \caption{Profiles of gas pressure and turbulent pressure, from two CIE hydrodynamic simulations with different metallicities (blue for {\tt Z0.1} and orange for {\tt Z1}) at $t=50 \ \mathrm{Myr}$: dashed line --- gas pressure $P$, solid line --- the sum of gas pressure and turbulent pressure $P + \langle \rho \delta v_z^2\rangle$ where $\delta v_z$ is the velocity dispersion along $z$ direction. The drop in gas pressure due to fast radiative cooling in the mixing layer is roughly compensated by turbulent pressure.}
      \label{fig:P_vy}
    \end{center}
\end{figure}

To understand the structure of the mixing layer in more detail, one must employ the fluid equations. First, it is useful to verify the scaling relations we have obtained in \S\ref{sec:OOM} from an order of magnitude analysis of the fluid equations. The mass conservation equation is used in the definition of the parameter $\alpha$ (Eq. \eqref{eq:alpha_define}), and we have discussed the energy conservation equation above. The streamwise momentum equation, Eq. \eqref{eq:vt_scaling} $v_\mathrm{t}^2\sim \alpha v_x^2$, gives the scaling between turbulent velocity and shear velocity. We have verified directly that if we use the value of $\alpha$ obtained from our simulations, this relation approximately holds. In addition, we have directly verified that Eq. \eqref{eq:pressure_const}, which comes from the transverse momentum equation and expresses the continuity of the sum of thermal and turbulent pressure, also approximately holds in our simulations (see Fig. \ref{fig:P_vy}).  

\begin{figure}
    \begin{center}
      \includegraphics[width=0.5\textwidth]{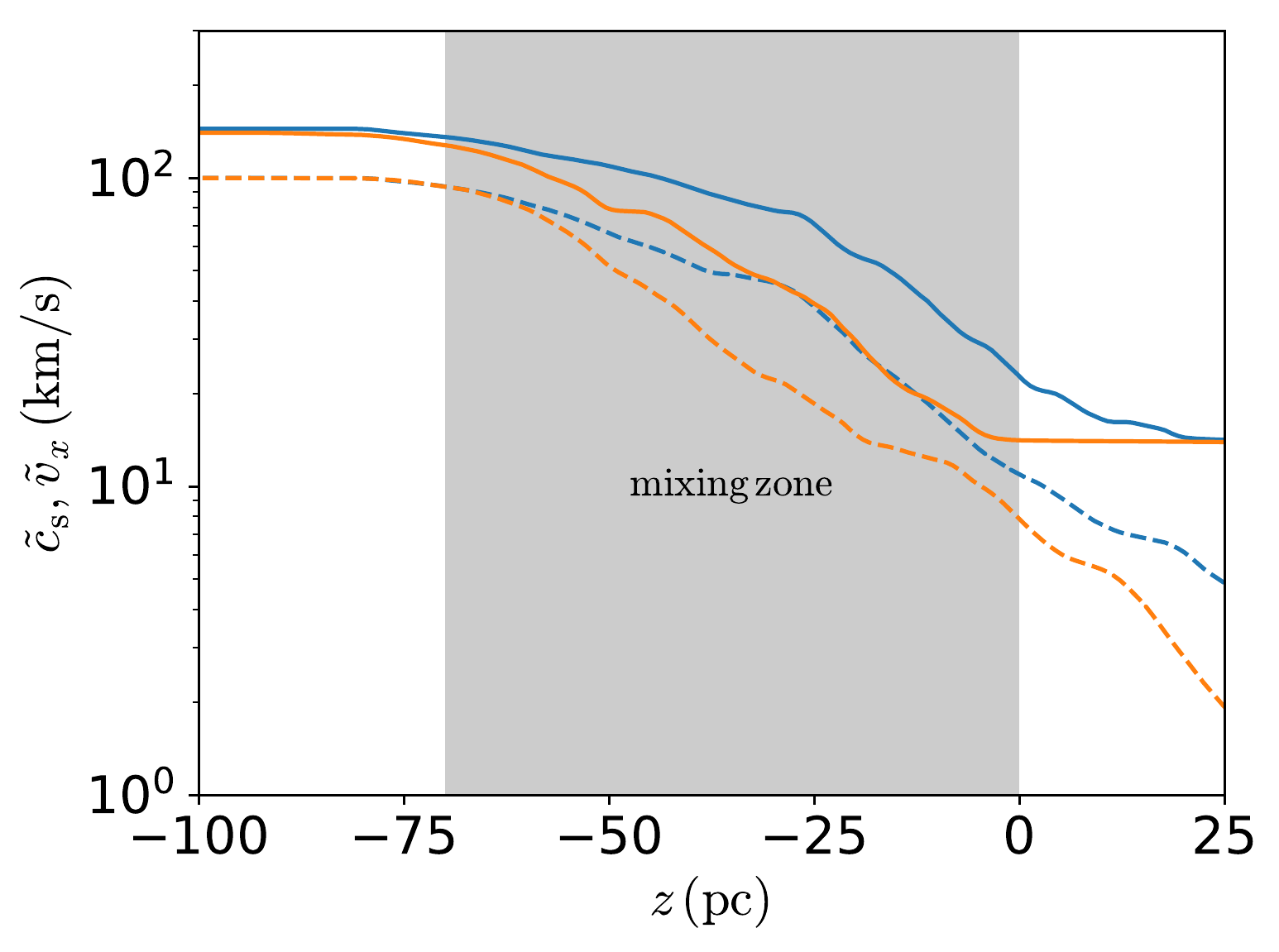}
      \caption{Profiles of gas sound speed and $x$-velocity, from two CIE hydro simulations with different metallicities (blue for {\tt Z0.1} and orange for {\tt Z1}) at $t=50 \ \mathrm{Myr}$: solid line --- sound speed $\tilde{c}_\mathrm{s}$, dashed line --- $x$-velocity $\tilde{v}_x$.  The approximate mixing zone is shadowed in gray in this figure. The shear velocity closely tracks the sound speed cross the mixing layer.}
      \label{fig:cs_vx}
    \end{center}
  \end{figure}

The order of magnitude analysis approach therefore appears reliable, and significantly simplifies calculations. However, as previously noted, the fluid equations are not closed, as we have 5 variables and 4 equations. Fortunately, one variable can be eliminated. An interesting feature in our simulations is that the average shear velocity $\tilde{v}_x(y)$ closely tracks the sound speed $\tilde{c}_\mathrm{s}(y)$ (see Fig. \ref{fig:cs_vx}). The Mach number of the flow is approximately constant across the mixing layer $\mathcal{M} \sim 0.6$, even when $\tilde{v}_x(y)$, $\tilde{c}_\mathrm{s}(y)$ vary by an order of magnitude. Thus, the ratio of kinetic to thermal energy density is approximately constant across the flow, implying efficient dissipation of bulk kinetic energy (e.g. via shocks) in radiative mixing layers. This implies that energy and momentum diffusion across the mixing layer track one another, and thus that the Prantl number of the flow is approximately constant. The ansatz $\tilde{v}_x(y) \propto c_\mathrm{s}(y)\propto \sqrt{T}$ allows us to eliminate one variable from our equations, resulting in $4$ equations in $4$ unknowns.

Given these facts, one could integrate the fluid equations to obtain mixing layer profiles. A simpler way to understand the structure of the mixing layer is via mixing length theory. In particular, if we assume that turbulent heat diffusion balances cooling, then: 
\begin{align}
  \nabla \cdot \left[\kappa_\mathrm{mix} \rho T \nabla s \right] = n_\mathrm{e} n_\mathrm{i} \Lambda(T),
  \label{eq:mixing_length}
\end{align}
where the specific entropy $s = c_\mathrm{P} \mathrm{ln}(P \rho^{-\gamma})$, $c_\mathrm{P} = 5/2\ k_\mathrm{B} / n m_\mathrm{H}$ is the specific heat, and $\kappa_\mathrm{mix}$ is the (unknown) mixing length diffusion coefficient. We need an ansatz for the latter. From Eq. \eqref{eq:mixing_length}, a reasonable choice is: 
\begin{align}
\kappa_{\rm mix} \sim \frac{\Sigma_{\rm L} h}{5/2 P} \sim 10^{26} \, {\rm cm^{2} \, s^{-1}}
\label{eq:kappa}
\end{align} 
Note that Eq. \eqref{eq:mixing_length} only applies in regions which are actively cooling and where thermal balance applies. This does reasonably well in our fiducial and low cooling cases, but less well in the high cooling case, where thermal balance fails at fairly high $T_{\rm crit}$. In any case, since the single phase approximation is a poor one for mixing layers (i.e., characterizing gas at fixed $z$ with a single density and temperature is not a good assumption), we do not attempt more elaborate models. 

Finally, as a useful point of reference, we find that \OVI column densities can be reasonably well fit by: 
\begin{equation}
N_\mathrm{OVI} \approx 3 \times 10^{12} \, {\rm cm^{-2}} \left( \frac{Z}{Z_{\odot}} \right)^{0.8} v_{100}^{1/4} P_{-13.7}^{1/2} 
\end{equation}
where the cooling curve dependence is now incorporated into the metallicity dependence. If considered separately, then $N_\mathrm{OVI} \propto [\Lambda(T)]^{-0.45}$, $N_\mathrm{CIV} \propto [\Lambda(T)]^{-0.53}$ and $N_\mathrm{SiIV} \propto [\Lambda(T)]^{-0.55}$, which is consistent with Eq. \eqref{eq:width} since $N_\mathrm{ion}\propto h \propto [\Lambda(T)]^{-1/2}$.

  \begin{figure*}
    \begin{center}
      \includegraphics[width=\textwidth]{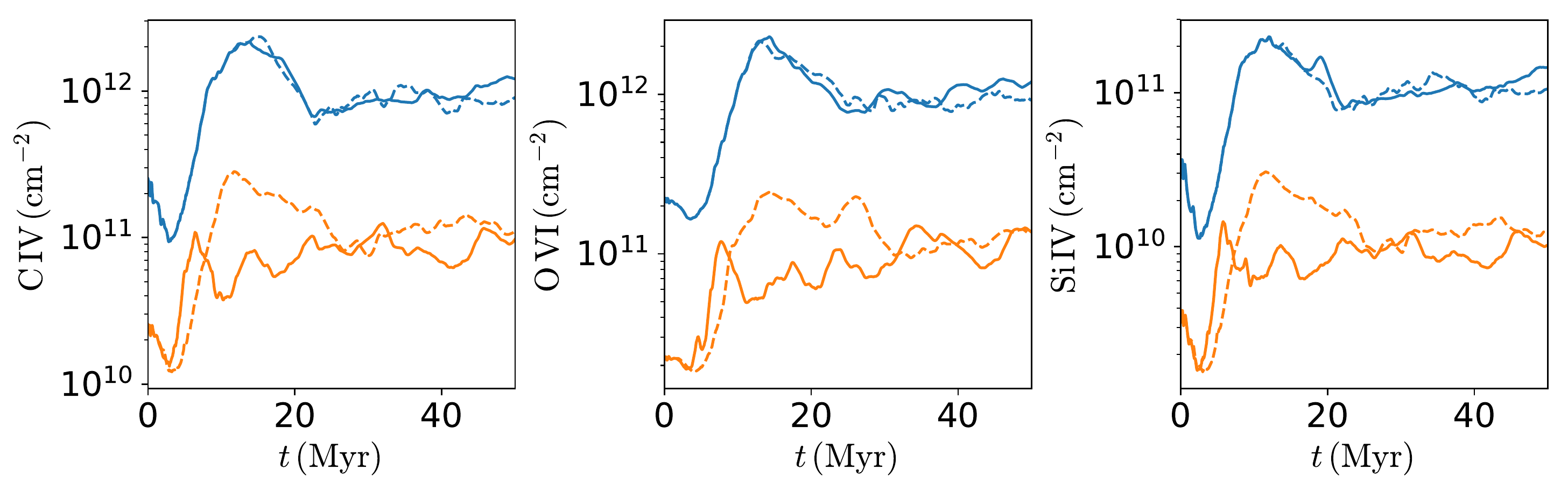}
      \caption{Effect of box width and perturbation wavelength on ion column densities: dashed blue --- {\tt Z1}, solid blue --- {\tt Z1\_size-lgr}, dashed orange --- {\tt Z0.1}, solid orange --- {\tt Z0.1\_wave-sml}.}
      \label{fig:wide_pertlen}
    \end{center}
  \end{figure*}

\subsection{MHD simulations}

We now turn to the effect of magnetic fields. Before we begin, it is useful to remind ourselves of results from the adiabatic MHD Kelvin Helmholtz instability, both from linear theory \citep{chandrasekhar61,miura82} and numerical simulations \citep{jones97,ryu00}. The geometry and amplitude of the initial field is important. For magnetic fields orthogonal to both the flow direction and interface normal (the $y$-direction in our simulations), tension does not stabilize the interface and the flow is essentially hydrodynamic\footnote{Of course, fields parallel to the interface normal strongly suppress shear, and there is no background equilibrium state.}. For fields parallel to the flow, the flow is either linearly stable ($\mathcal{M}_{\rm A} \lsim 2$), quickly stabilized by modest growth of corrugations in the perturbed shear layer ($2 \lsim \mathcal{M}_{{\rm A}x} \lsim 4$), or initially hydrodynamic but eventually stabilized by magnetic fields which are amplified by vortex stretching and twisting ($4 \lsim \mathcal{M}_{{\rm A}x} \lsim 50$). For $\mathcal{M}_{{\rm A}x} \gsim 50$, dissipation due to magnetic reconnection appears to render the flow hydrodynamic to the end, at least in $256^{3}$, 3D simulations \citep{ryu00}.

\begin{figure*}
    \begin{subfigure}[b]{\textwidth}
      \includegraphics[width=0.910\textwidth]{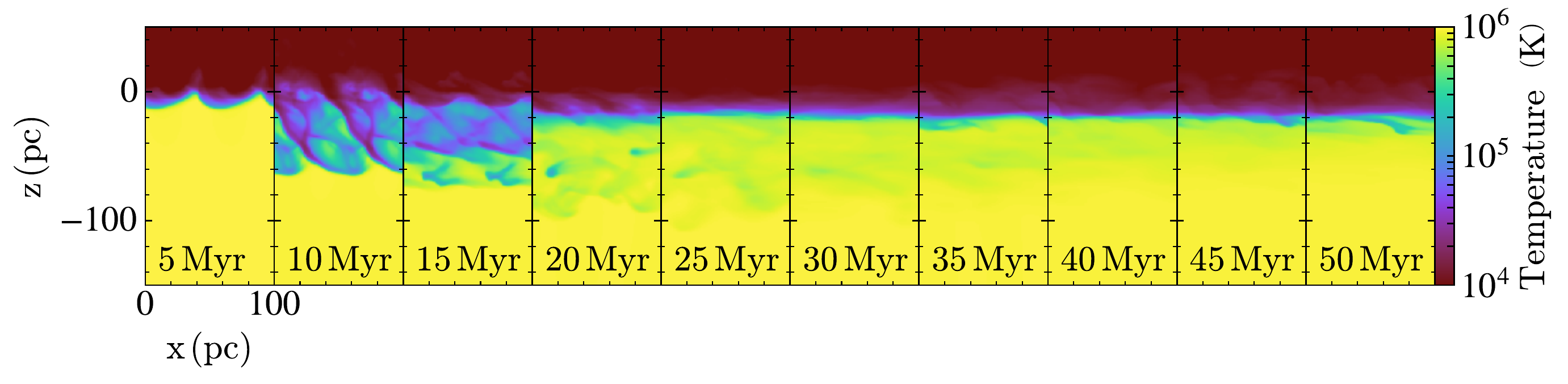}
      \includegraphics[width=0.918\textwidth]{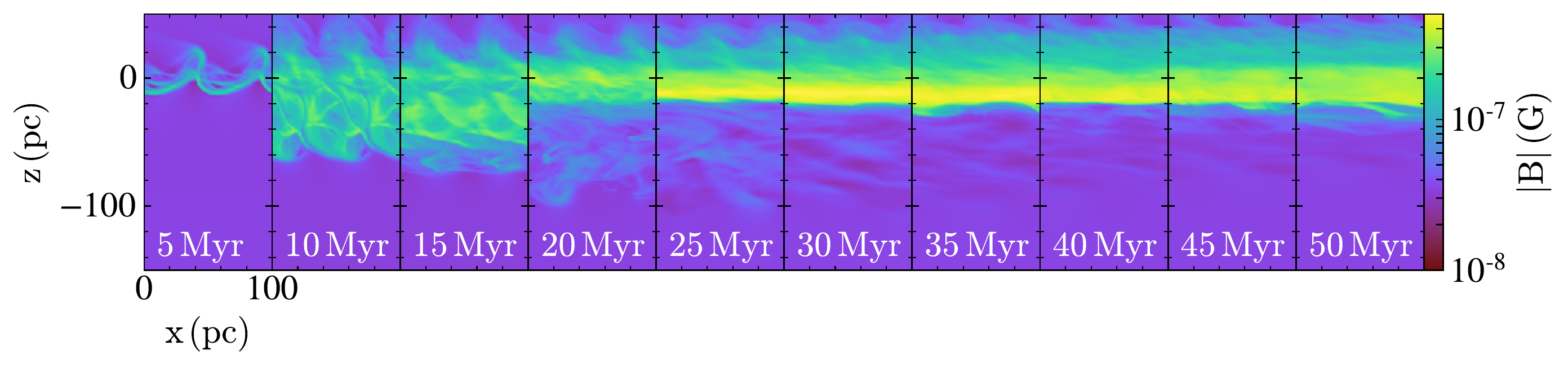}
      \includegraphics[width=0.913\textwidth]{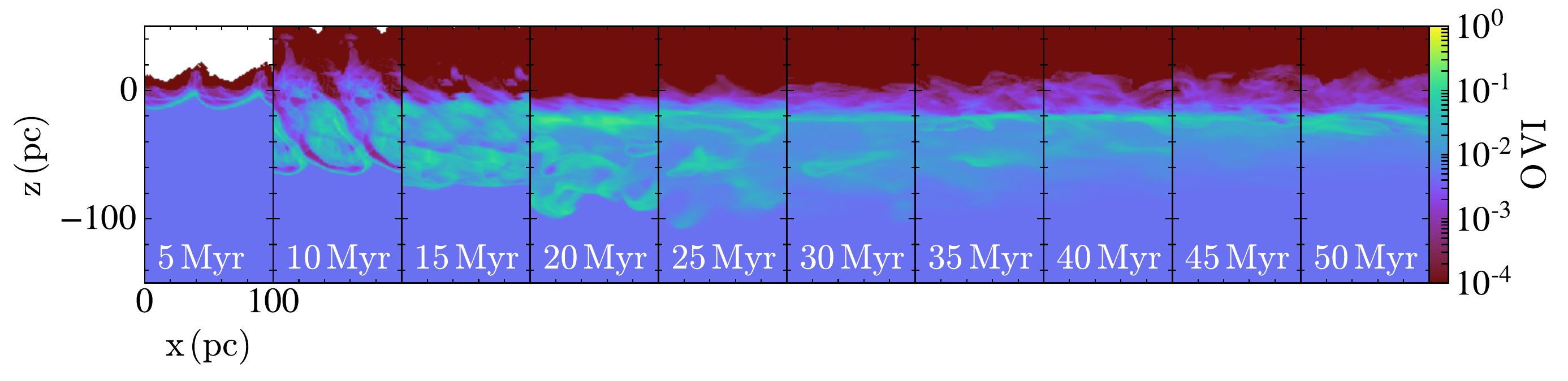}
      \caption{\tt Z1\_Bx-sml}
      \label{fig:cie_Bx-sml}
    \end{subfigure}
    \begin{subfigure}[b]{\textwidth}
      \includegraphics[width=0.915\textwidth]{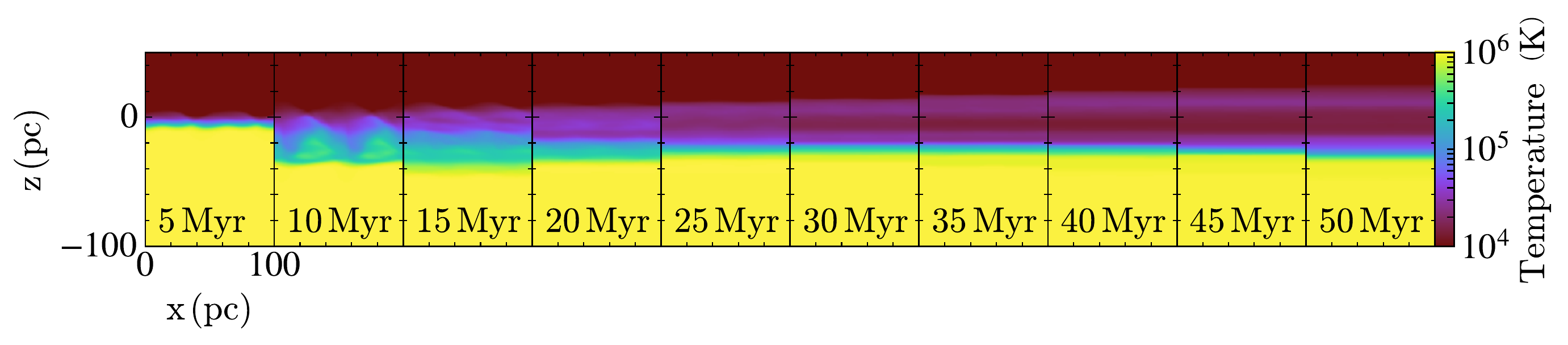}
      \includegraphics[width=0.95\textwidth]{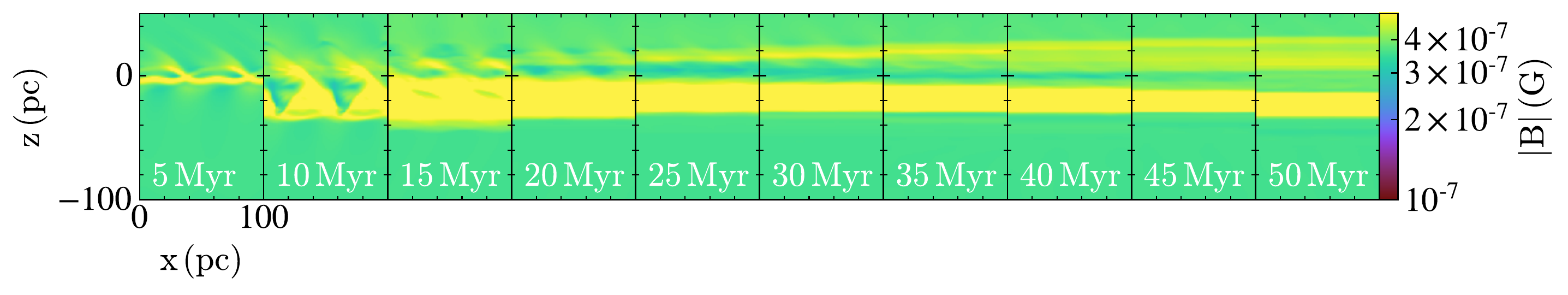}
      \includegraphics[width=0.915\textwidth]{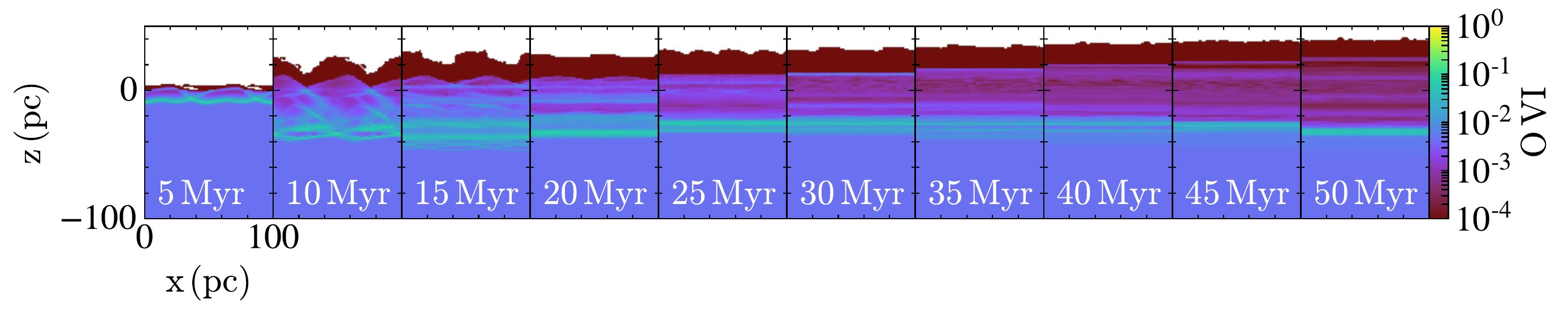}
      \caption{\tt Z1\_Bx-lgr}
      \label{fig:cie_Bx-lgr}
    \end{subfigure}
    \caption{The timeseries of density-weighted projections of temperature, \OVI fraction and magnetic field strength, from CIE MHD simulations (a) {\tt Z1\_Bx-sml} and (b) {\tt Z1\_Bx-lgr}. In both cases, mixing is strongly suppressed by the end of the simulation.}
    \label{fig:proj_cie_Bx}
  \end{figure*}

With some small changes, we find similar results in MHD simulations with radiative cooling. In Fig. \ref{fig:proj_cie_Bx} we present the projection plots from MHD simulations with initially horizontal fields --- runs {\tt Z1\_Bx-sml} and {\tt Z1\_Bx-lgr} with initial $\beta$ of $430$ ($\mathcal{M}_{\rm A} = 13$) and $4.3$ ($\mathcal{M}_{\rm A} = 1.3$), showing the time evolution of temperature, magnetic field strength and \OVI fraction.

	For the weak field simulation (Fig. \ref{fig:cie_Bx-sml}), the evolution of temperature and \OVI fraction during the first $15 \ \mathrm{Myr}$ is very similar to that in the hydrodynamic simulation (Fig. \ref{fig:proj_CIE}). However, from $\sim 20 \ \mathrm{Myr}$, the region with warm temperature and relative high \OVI fraction starts to shrink. The projection plot of magnetic field strength suggests that during the first $15 \ \mathrm{Myr}$, magnetic fields are amplified significantly and quickly spread over entire turbulent region. Starting from $25 \ \mathrm{Myr}$, strong magnetic fields develop at the interface, with the strength growing to $\sim 10$ times the initial value. The results are consistent with field amplification until the magnetic energy density reaches equipartition with the turbulent energy density, and the flow reaches $\mathcal{M}_{\rm A} \sim 1$. The amplification is highest at the cold hot interface, where most of the stretching due to turbulence takes place. At this point, mixing is strongly suppressed by magnetic tension, and the flow becomes laminar. The evolution of the strong field case is shown in Fig. \ref{fig:cie_Bx-lgr}. Compared to the weak field case, turbulence at the interface is significantly suppressed from the very beginning, and quickly transforms into laminar flow. Magnetic fields are also amplified, but by a relatively small factor of $\sim 2$ compared with the weak field case. With initially strong fields, the turbulent mixing layer is much less developed at all times. 

  \begin{figure*}
    \begin{center}
      \includegraphics[width=\textwidth]{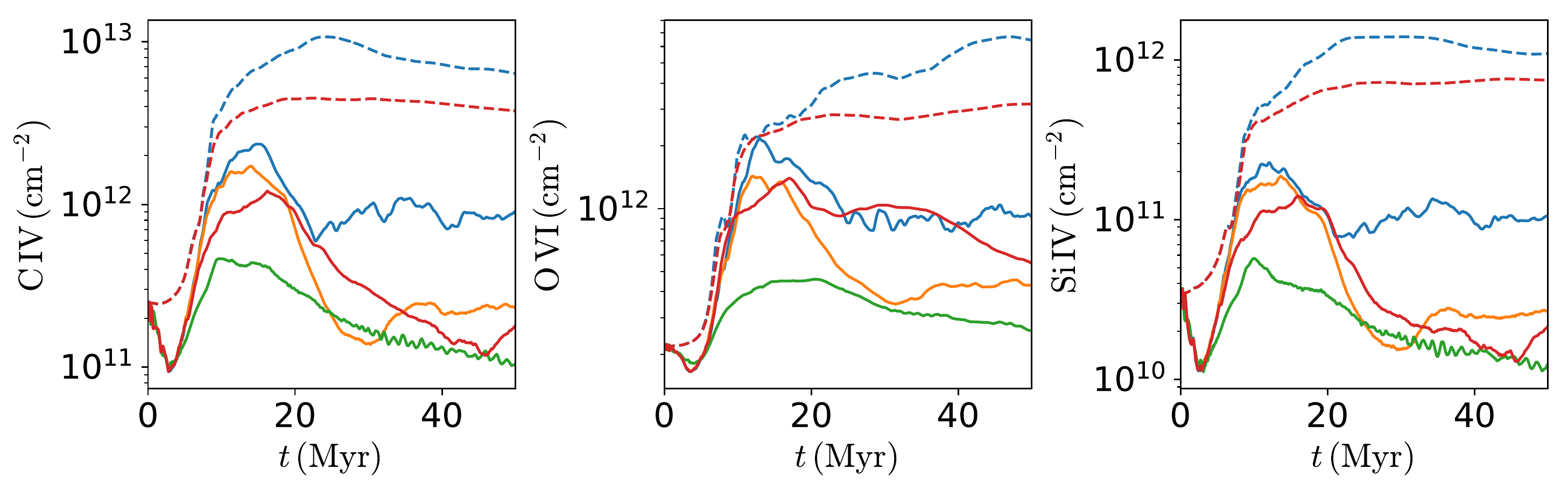}
      \caption{Effect of horizontal ($x$-axis-aligned) and transverse ($y$-axis-aligned) magnetic fields magnetic fields on ion column densities. Dashed lines indicate simulations without radiative cooling. Solid blue --- {\tt Z1}, dashed blue --- {\tt Z1\_cool0}, orange --- {\tt Z1\_Bx-sml}, green --- {\tt Z1\_Bx-lgr}, solid red --- {\tt Z1\_By-sml}, dashed red --- {\tt Z1\_By-sml\_cool0}. Regardless of initial strength or geometry, magnetic fields strongly suppress mixing and the contribution of the mixing layer to ionic columns.}
      \label{fig:bxbz}
    \end{center}
  \end{figure*}

The time evolution of ion column density from MHD simulations also suggests a similar trend. In Fig. \ref{fig:bxbz}, during the first $\sim 10 \ \mathrm{Myr}$, the column density in the weak field case (orange) grows exponentially at a similar rate with that in hydro case, and also peaks at similar but slightly lower levels. However, the weak field case starts to differ from the hydrodynamic case afterwards: instead of becoming stable at a certain level, the ion column density exponentially decreases. The column density in strong field case (green) differs from either hydrodynamic or weak field case from the very beginning: the growth rate is slower, the peak appears at a much smaller value, and the ion column density decays with a slower rate than the weak field case. In either case, regardless of initial field strength, the final column densities are roughly the same as their initial value: comparable or less than initial columns, indicating that mixing layers make no contribution to ion column densities, as one would expect for laminar flow. 

The effects of transverse fields ($y$-axis-aligned, where field lines are parallel with the mixing plane but perpendicular to the interface normal) can also be seen on this plot (red curves). The transverse case without cooling more or less coincides with the hydrodynamic case without cooling, as expected from previous studies (red vs. blue dashed lines). However, the transverse case with radiative cooling does {\it not} behave like the radiative hydrodynamic case; column densities drop drop after an initial increase, similar to the MHD case with flow oriented field lines in the $x$-direction. We suspect this is because radiative cooling causes small scale motions which perturb the field direction, introducing components which are eventually amplified to stabilize the mixing layer. 

In short: magnetic fields in a mixing layer can always develop and shut down turbulent mixing, \emph{independent} of initial field strength and geometry. Note that in the MHD case, the driving of turbulence by radiative cooling is weaker, since there is less mixing and hence less cooling. Low thermal pressures which develop from fast cooling can also be compensated by magnetic pressure (which is most strongly amplified in the mixing layer) rather than turbulent pressure. While the turbulence eventually decays once mixing (and thus radiative cooling) is suppressed, magnetic support will remain. 

\subsection{Convergence}

\begin{figure*}
    \begin{center}
      \includegraphics[width=\textwidth]{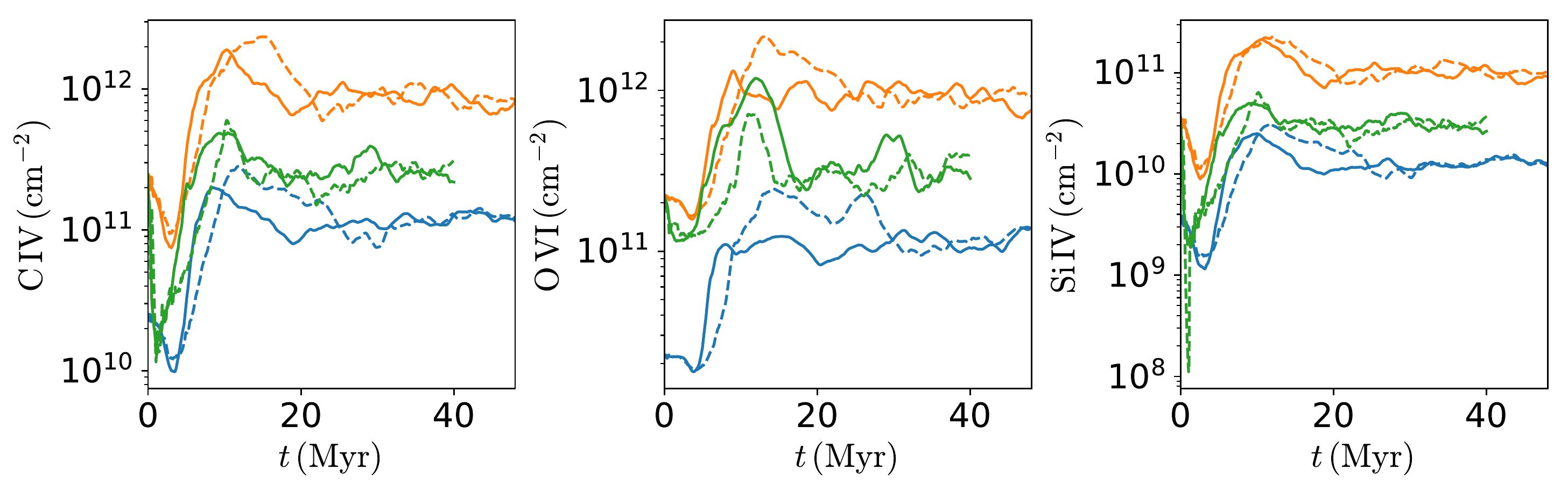}
      \caption{Convergence test for hydro simulations with different metallicities and cooling rates: dashed blue --- {\tt Z0.1}, solid blue --- {\tt Z0.1\_reso-lgr}; dashed orange --- {\tt Z1}, solid orange --- {\tt Z1\_reso-lgr}; dashed green --- {\tt Z1\_cool10}, solid green --- {\tt \tt Z1\_cool10\_reso-lgr}. Our simulations are converged. }
      \label{fig:conv_hydro}
    \end{center}
  \end{figure*}

\begin{figure*}
    \begin{center}
      \includegraphics[width=\textwidth]{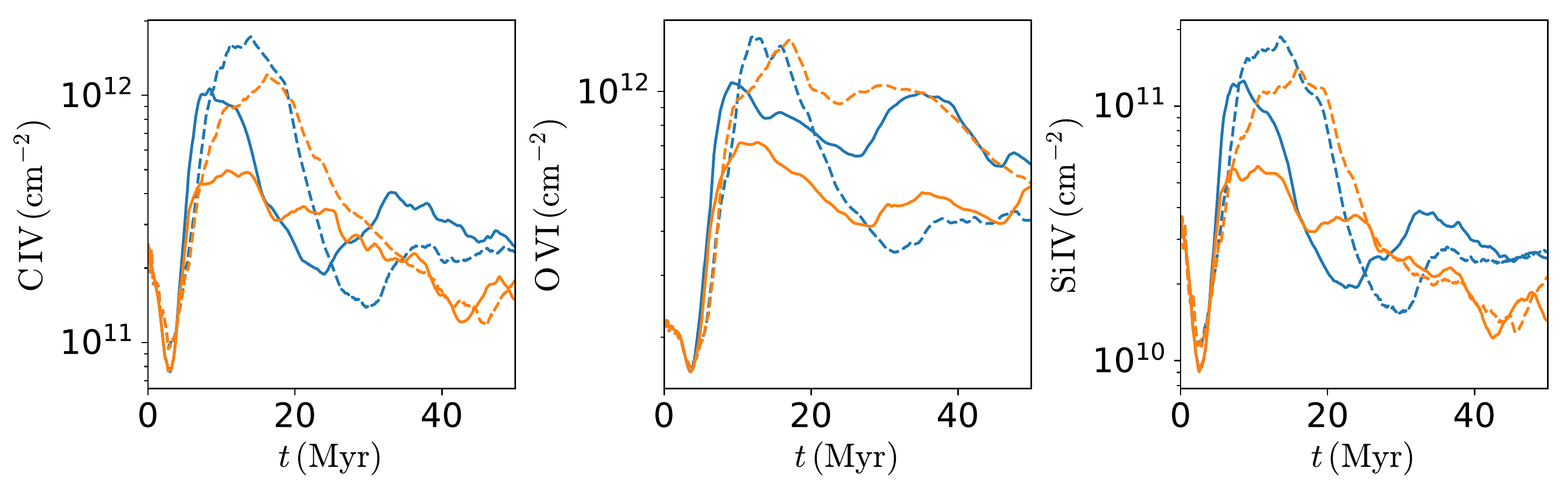}
      \caption{Convergence test for MHD simulations with different metallicities and cooling rates: dashed blue --- {\tt Z1\_Bx-sml}, solid blue --- {\tt Z1\_Bx-sml\_reso-lgr}; dashed orange --- {\tt Z1\_By-sml}, solid orange --- {\tt Z1\_By-sml\_reso-lgr}. Higher resolutions show a lower initial peak due to faster magnetic field growth, which limits mixing earlier. However, the final steady state column densities are roughly converged.}
      \label{fig:conv_mhd}
    \end{center}
  \end{figure*}

  We present our convergence tests for the hydro case in Fig. \ref{fig:conv_hydro} and the MHD case in Fig. \ref{fig:conv_mhd}. For hydro simulations, the ion column densities are well converged, even for the strong cooling case. Instead, for the MHD case, although the final level of ion column densities for fiducial and high-resolution runs are similar, the peaks of column densities in high resolution runs are significantly lower than those in low resolution runs. This is due to the fact that vorticity under higher resolution is better resolved and thus reaches a larger magnitude, which results in faster amplification of magnetic fields, and thus earlier suppression of mixing layers which limits the growth of ion column densities. Thus, the pattern of growth followed by decay is partly a resolution effect which may be much less pronounced at high resolution. However, as long as we can accurately calculate the final steady state, we are less concerned by differences in transient behavior. 
  
\subsection{Photoionization}

\begin{figure*}
    \begin{center}
      \includegraphics[width=\textwidth]{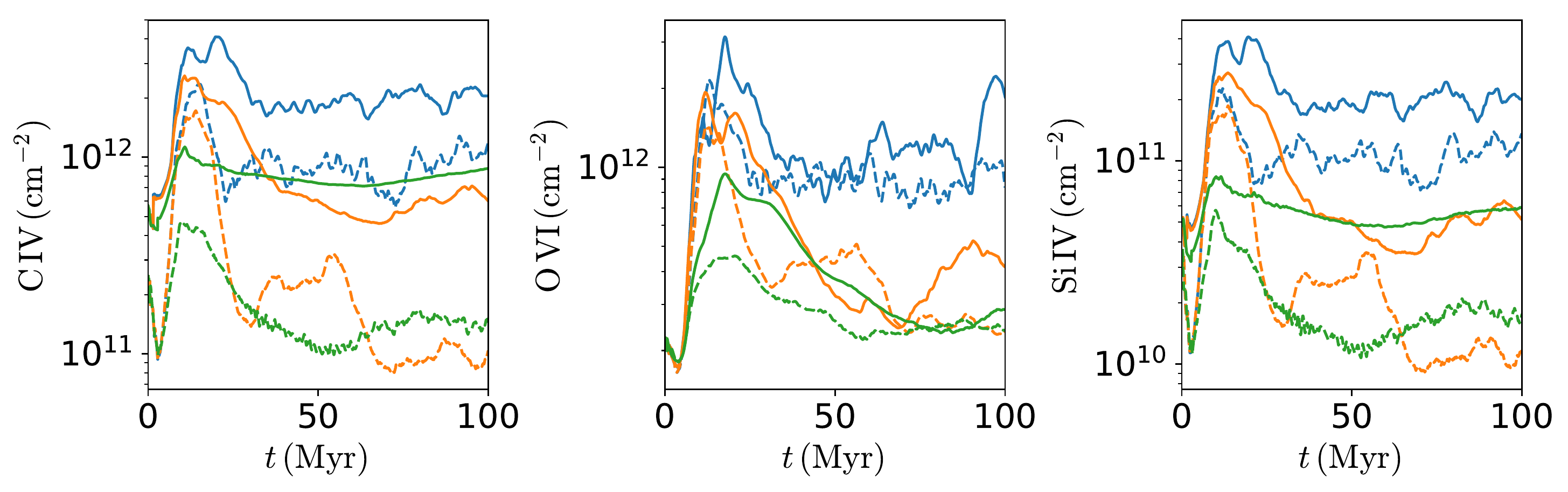}
      \caption{Effect of photoionization on ion column densities: dashed blue --- {\tt Z1}, solid blue --- {\tt Z1\_photo}; dashed orange --- {\tt Z1\_Bx-sml}, solid orange --- {\tt Z1\_Bx-sml\_photo}; dashed green --- {\tt Z1\_Bx-lgr}, solid green --- {\tt Z1\_Bx-lgr\_photo}.}
      \label{fig:eqi_ph_noph}
    \end{center}
  \end{figure*}

To investigate the importance of photoionization, we perform runs {\tt Z1\_photo}, {\tt Z1\_Bx-sml\_photo} and {\tt Z1\_Bx-lgr\_photo}, which are shown in Fig. \ref{fig:eqi_ph_noph}. Generally speaking, photoionization can affect the simulations in two ways. First, ions can be photoionized; second, photoionization can modify hydrodynamics by altering the cooling curves, which is shown in Fig. \ref{fig:cooling}. The first effect is dominated primarily by gas outside the mixing layer (whose amount we do not model); thus, the second effect is of most interest to us. The first effect shows up immediately in the initial conditions: the photoionized case (solid lines) shows large initial column densities in \CIV, \SiIV compared to the CIE case\footnote{For the initial conditions we have chosen, with the UV radiation background we assumed, \OVI cannot be significantly photoionized: it is collisionally ionized in $T\sim 10^{6}$ K gas, and the recombination rate is too high in the dense $T \sim 10^{4}$ K gas. However, our poor knowledge of the amplitude of the EUV band of the metagalactic background creates uncertainty in  the true photoionization rate of \OVI.} (dashed lines). Thus, to tease out the effect of photoionization on mixing layers, we should look at the difference between initial and final column densities, rather than comparing absolute column densities. 
  
By comparing the solid and dashed curves, (with and without photoionization respectively), we see that photoionization has relatively little effect on the mixing layer. Photoionization does boost the initial column density of \SiIV and \CIV, primarily by photoionizing gas outside the mixing layer (where there is a lot more mass). However, in the MHD runs, after a transient, the final column density returns to roughly the initial column density for the photoionized case, just as for the CIE case. Thus, the same conclusions apply -- as mixing is strongly suppressed in the MHD case, there is little ionic column at the interface between different gas phases. In the hydrodynamic case, where there is significant mixing and thus increase in ionic column, the photoionized and CIE cases differ by at most a factor of two for \CIV, \SiIV, and even less for \OVI. Similar conclusions hold for the lower pressure case ($nT = 10 \, {\rm cm^{-3} K}$, not shown), where one might have expected photoionization to be more important. Thus, in the solar metallicity case, photoionization does not play an important role. Interestingly, we have found that photoionization {\it does} boost \OVI column density in the $Z=0.1 \, Z_{\odot}$ case, particularly with MHD, where the column density is a factor of $\sim 5$ larger than in the CIE case (not shown; see further discussion in \S\ref{sect:NEI}).
   
We can understand these results as follows. As discussed above, the main effect of photoionization on radiative mixing layers is to change the radiative efficiency: ions are ``overionized'' for the plasma temperature, making collisional excitation and cooling less efficient. For pressures of $nT=100, \, 10 \, {\rm cm^{-3} \, K}$, where $T\sim 10^{5}$ K gas has densities of $n \sim 10^{-3}, 10^{-4} \, {\rm cm^{-3}}$, at solar metallicity the cooling curve does not change significantly (Fig. \ref{fig:cooling}). However, for $Z= 0.1 \, Z_{\odot}$, the change in the cooling curve {\it is} significant for this range of densities, particularly in the $T \sim$ few $\times 10^{4}\ \mathrm{K}$ range. This is because the main coolants in this temperature range, hydrogen and helium, are easily photoionized by the background radiation field. Long cooling times increase the column density of mixing layers; they also reduce the enthalpy flux from hot gas required to offset cooling, thus allowing significant column density in the MHD case. By contrast, at solar metallicity, other metal ions can still perform the lion's share of cooling, with little change in the cooling curve. Similar effects arise in time-dependent (NEI) calculations of cooling with photoionization (see Fig 4 of \citealt{gnat17}).

\subsection{Equilibrium vs. non-equilibrium ionization}
\label{sect:NEI}

Previous studies have shown that non-equilibrium ionization (NEI) boosts column densities of \CIV, \OVI, \SiIV in mixing layers by a factor of a few \citep{kwak10}. We reproduce this effect in our hydrodynamic simulations, and find from examining tracer particles that it arises due to the fact that due to mixing, these ions continually bounce between warm and hot gas on timescales shorter than the ionization/recombination time. Here we examine the impact of magnetic fields and photoionization on NEI effects. 

\begin{figure*}
    \begin{center}
      \includegraphics[width=\textwidth]{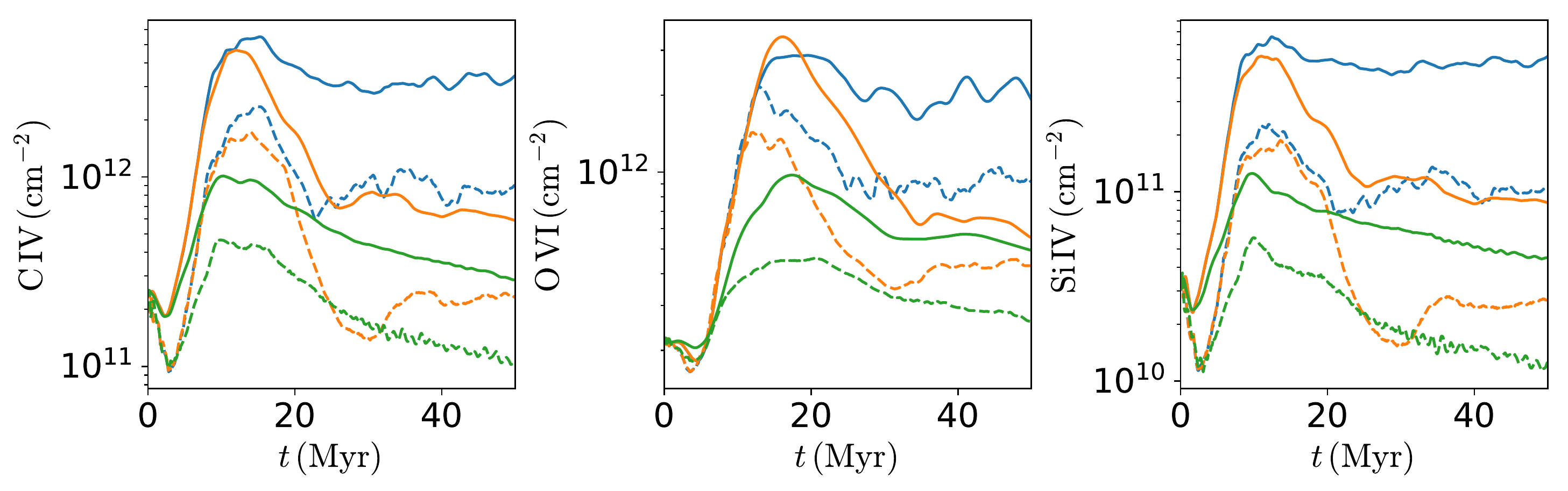}
      \caption{Effect of non-equilibrium ionization on ion column densities: dashed blue --- {\tt Z1}, solid blue --- {\tt Z1\_nei}; dashed orange --- {\tt Z1\_Bx-sml}, solid orange --- {\tt Z1\_Bx-sml\_nei}; dashed green --- {\tt Z1\_Bx-lgr}, solid green --- {\tt Z1\_Bx-lgr\_nei}.}
      \label{fig:eqi_nei}
    \end{center}
  \end{figure*}

\begin{figure*}
    \begin{center}
      \includegraphics[width=\textwidth]{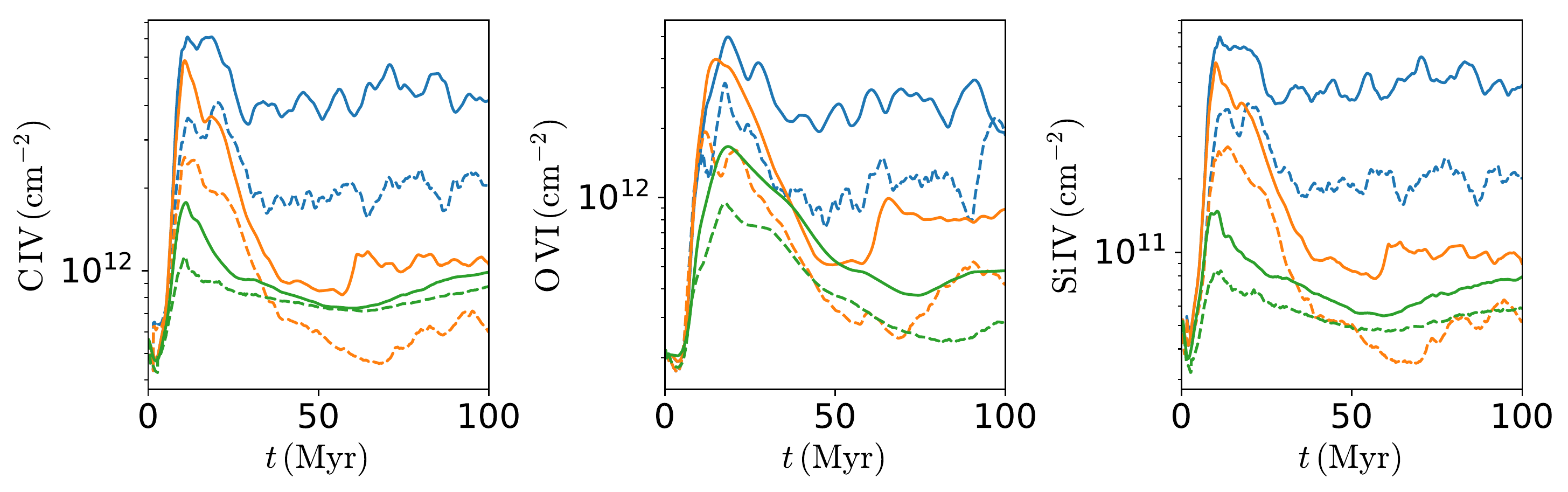}
      \caption{Effect of the coexistence of non-equilibrium ionization and photoionization on ion column densities: dashed blue --- {\tt Z1\_photo}, solid blue --- {\tt Z1\_nei\_photo}; dashed orange --- {\tt Z1\_Bx-sml\_photo}, solid orange --- {\tt Z1\_Bx-sml\_nei\_photo}; dashed green --- {\tt Z1\_Bx-lgr\_photo}, solid green --- {\tt Z1\_Bx-lgr\_nei\_photo}.}
      \label{fig:nei_ph}
    \end{center}
  \end{figure*}

\begin{figure}
    \begin{center}
      \includegraphics[width=0.5\textwidth]{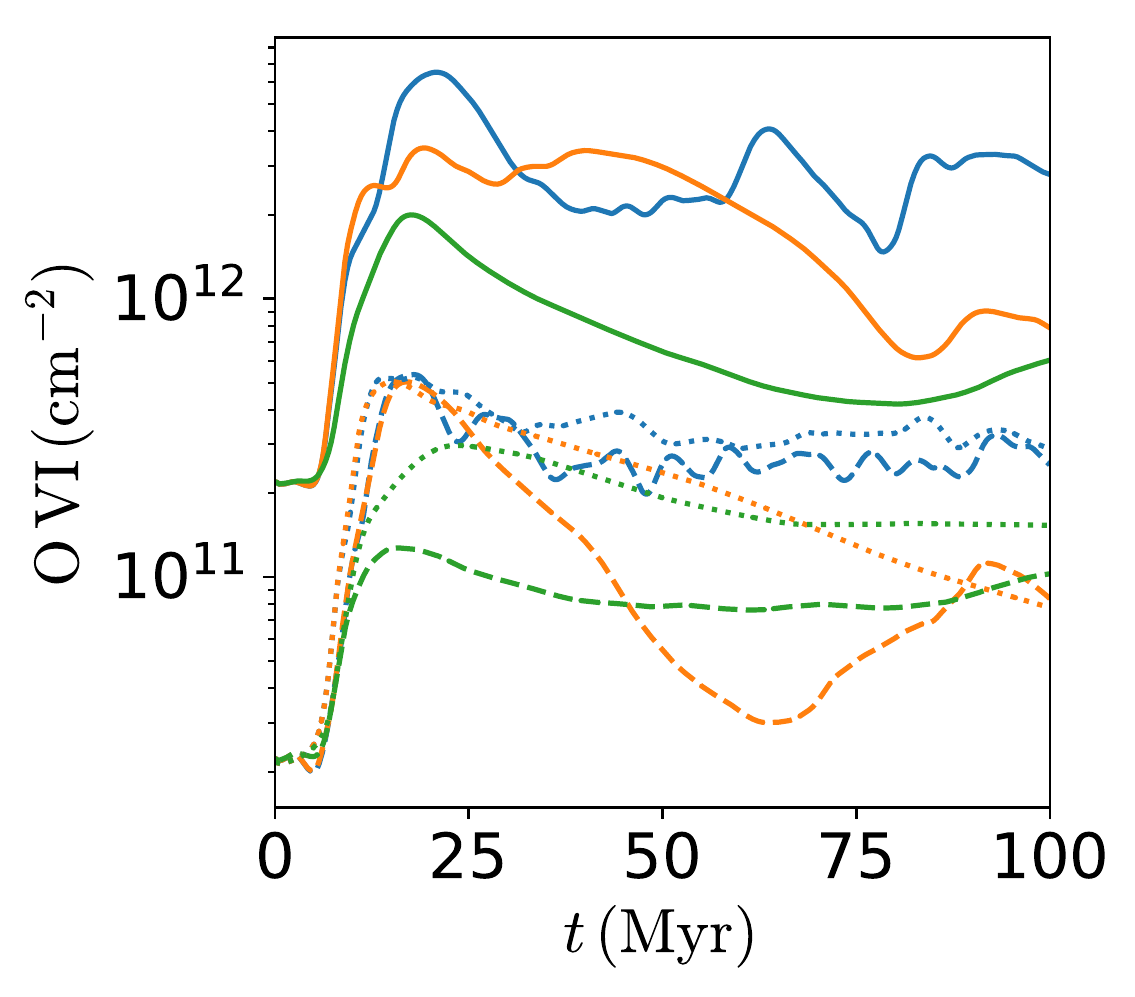}
      \caption{Effect of time-dependent cooling curve, photoionization and NEI on \OVI column densities with $1$ and $0.1$ solar metallicities and different magnetic field strengths: solid lines --- {\tt Z1\_nei\_cool-timedp\_}, dashed lines --- {\tt Z0.1\_nei\_}, dotted lines --- {\tt Z0.1\_nei\_photo\_}; blue --- hydro, orange --- {\tt \_Bx-sml}, green --- {\tt \_Bx-lgr}.}
      \label{fig:nei_ph_cool}
    \end{center}
  \end{figure}

In Fig. \ref{fig:eqi_nei}, we show NEI (solid) and CIE (dashed) calculations for the hydrodynamic and MHD cases (both strong and weak initial fields). In all cases, the NEI column densities are indeed larger than the CIE column densities by a factor of $\sim 1.5-3$. Thus, in the MHD case, where in CIE mixing layers do not contribute appreciably to low ion columns of \SiIV, \CIV (initial and final dashed values are similar), once NEI effects are taken into account, mixing layers now contribute appreciably. For \OVI, where mixing layers {\it do} contribute even in the CIE case, a long lasting boost by a factor of $\sim 2$ is also evident. Similar column density boosts of a factor of $\sim 1.5-3$ are also seen in the full case where both photoionization and NEI effects are taken into account (Fig. \ref{fig:nei_ph}). Indeed, at solar metallicity, photoionization has relatively little effect on NEI effects (compare solid lines in Fig. \ref{fig:eqi_nei} \& \ref{fig:nei_ph}); one would have to go to lower pressures to see the impact of photoionization. The same is true at $Z=0.1 Z_{\odot}$ (Fig. \ref{fig:nei_ph_cool}; compare dashed and dotted lines), even though we previously remarked that in the CEI case, photoionization provided a column density boost in the low metallicity case. If we compare the low-metallicity photoionized (not shown), NEI, and NEI+photoionized case, they all show roughly the same boost over the CIE case; it appears that these effects do not compound.  

One effect we have not yet modeled is the effect of non-equilibrium ionization on cooling. Similar to photoionization, NEI effects reduce cooling efficiencies, since ions are over-ionized for a given plasma temperature. As previously noted, we do not fully self-consistently track how the cooling function changes due to NEI effects, but simply use the CIE (or PIE) cooling function. However, one way to model this is to utilize calculations of time-dependent isobaric collisional cooling which do compute the NEI cooling curve \cite{gnat07}; NEI effects appear because the cooling time can be shorter than the recombination time. This does not capture the additional contribution of mixing to NEI effects (i.e., if the mixing time is shorter than the recombination time), but can serve as a useful guage. The result is shown in Fig. \ref{fig:nei_ph_cool} (solid curves; compare to solid curves in Fig. \ref{fig:eqi_nei}). At least for solar metallicity, NEI effects on the cooling curve do not produce a significant change in column densities; the change in the cooling curve is too small.

\section{Conclusions}
\label{sect:conclusions}

Turbulent mixing layers (TMLs) should be ubiquitous in astrophysics, arising wherever there is shear at the boundary between different gas phases. TMLs are an attractive explanation for the prevalence of high ions (e.g. \OVI) seen in high velocity clouds and in the circumgalactic medium of galaxies: they can explain the steady-state persistence of such ions despite the short cooling times of the $T\sim 10^5 \ \mathrm{K}$ gas which host them; they can explain the observed kinematic coincidence between low and high ions, and simple analytic estimates give column densities in rough agreement with observations. They are also potentially relevant for understanding the low to intermediate-velocity forbidden emission line components of protostellar outflows \citep{pyo2003adaptive,white2016alma}, and the growth and entrainment of cold clouds in a galactic wind \citep{gronke2018growth}. Surprisingly, the literature probing the non-linear saturated state of the radiative Kelvin-Helmholtz instability is relatively sparse, and there is no direct comparison of analytic theory \citep{begelman90} with numerical simulations. In this paper, we run 3D hydrodynamic simulations with with an eye toward comparing analytic predictions with simulations. We also explore the impact of non-equilibrium ionization (NEI), photoionization (which had previously been ignored), and magnetic fields (which previously had not been run long enough for a stable equilibrium to develop).  Our results are as follows:

\begin{enumerate}

  \item \emph{Analytic scalings don't work}

  Analytic models assume a mixing layer with thickness $\sim v_\mathrm{turb} t_\mathrm{cool}$, where the turbulent velocity is assumed to be of order the shear velocity $v_\mathrm{turb} \sim v_\mathrm{shear}$. Since $t_\mathrm{cool} \propto 1/Z n $ in the critical $T\sim 10^5 \ \mathrm{K}$ range, this implies column densities which are roughly independent of density and metallicity. By contrast, we find turbulent velocities which are much smaller (and roughly independent of) the shear velocity, leading to significantly lower column densities. We also find that column densities fall as metallicity and density decrease.

  \item \emph{Radiative cooling seeds turbulence and drives hot gas entrainment}

  The above results suggest that the dynamics of gas entrainment and mixing is not driven primarily by the Kelvin-Helmholtz instability. Indeed, in our parameter study, we find that entrainment velocity of the hot gas $v_z^\mathrm{hot}$ is roughly independent of the shear velocity and overdensity, which implies that the Kelvin-Helmholtz instability is not primarily responsible for drawing hot gas, which carries enthalpy and momentum, into the mixing layer.\footnote{By contrast, the rate of mass entrainment of cold gas, $\dot{m}_\mathrm{cold} \simeq \sqrt{\rho_\mathrm{cold}/\rho_\mathrm{hot}} \dot{m}_\mathrm{hot}$, is set by Kelvin-Helmholtz instability. Cold gas entrainment dominates the mass budget, whereas hot gas entrainment dominates the energy budget.} Instead, cavities in thermal pressure created by fast radiative cooling in mixed gas create turbulence and draw hot gas into the mixing layer; the sum of thermal and turbulent pressure $P+\rho v^2$ is roughly continuous across the layer. We see that as the radiative cooling rate increases, both the hot gas entrainment rate and the level of turbulence increases, whereas these quantities are stable to the imposed shear. The energy lost via radiative cooling is balanced by enthalpy flux from hot gas entrainment. Overall, we find the inflow velocity $\tilde{v}_{y}^{\rm hot} \propto t_{\rm cool}^{-1/2}$, so that the mixing layer width $h \propto \tilde{v}_{y}^{\rm hot} t_{\rm cool} \propto t_{\rm cool}^{1/2}$, rather than $h \propto t_{\rm cool}$. Turbulent dissipation can also provide a comparable contribution when cooling times are short and turbulence is vigorous. 

  \item \emph{Magnetic fields suppress mixing}

  We also find that, independent of initial field strength and geometry, magnetic fields can be amplified to close to equipartition to turbulence, at which point magnetic tension suppress further mixing and the flow becomes quasi-laminar. These results are in rough accord with expectations from MHD simulations of the adiabatic Kelvin-Helmholtz instability, although it is clear that radiative cooling alters the detailed structure of the mixing layer.

  \item  \emph{Effect of Photoionization}
   
Photoionization --- which has not been previously included in hydrodynamic simulations --- has two effects: a) it directly increases ionic column densities, b) it reduces cooling rates, both due to the effects of photoheating and by rendering gas overionized for a given temperature. The first effect is primarily important in low-density gas outside the mixing layer, and has been considered in detail by others. The second is more pertinent to our study, and has an interesting effect: by increasing the cooling time, the metagalactic UV ionizing background can potentially increase column densities. However, for CGM conditions, once NEI (see below) and magnetic fields are taken into account, this is largely negligible. Photoionization therefore appears to have little influence on mixing layer column densities, although it is obviously still important for gas outside the interface. 
 
\item \emph{Effects of non-equilibrium ionization}

  In agreement with previous work \citep{kwak10}, we find that NEI simulations have column densities a factor of few larger than CIE calculations, as the temperature of gas rich in high ions changes on timescales shorter than the recombination time, due both to radiative cooling and turbulent mixing. We show that this continues to be true once magnetic fields and photoionization are taken into account. Thus, NEI effects allow significant ion column densities even when magnetic fields suppress mixing. Cooling curve suppression by NEI effects does not appear to play a significant role at solar metallicity.

\end{enumerate}

 Other heating mechanisms such as anisotropic thermal conduction \citep{borkowski90,gnat10}, or field-aligned cosmic-ray streaming, which both heats and pressurizes the gas \citep{wiener17-cold-clouds}, could significantly alter the structure of mixing layer. \footnote{The level of thermal conduction relative to canonical Spitzer values is highly uncertain, in light of small scale field line wandering as well as whistler or firehose plasma instabilities which change the electron mean free path. We implicitly assume a low level of thermal conduction, required to bring mixed gas in a grid cell to a common temperature. In our simulations, it is simply given by numerical diffusion. In principle, we should include explicit thermal conduction to guarantee convergence, as is necessary in simulations of classical thermal instability \citep{koyama04}. In practice, we have found our simulations to be sufficiently converged, likely due to thermal diffusivity due to turbulence.} In particular, each of these now introduce characteristic scale lengths given by the Field length \citep{field65} and $L_{\rm c} \sim v_{\rm A}{P_{c}}{n^{2} \Lambda(T)}$ \citep{wiener17-cold-clouds} respectively, where $v_{\rm A}$ is the Alfv\'en velocity. They also modify the dynamics. For instance, in the static plane-parallel case with conduction, depending on gas pressure cold gas can either evaporate or condense. Both conduction and cosmic rays have been considered in isolation in the static case but not in conjunction with turbulent mixing. We plan to do so in the future. 
  
Overall, our results suggest that a sightline must pierce hundred or even thousands of mixing layers to explain observations, which is difficult to conceive in standard models, but may be possible of cold gas consists of a ``fog'' of dense, small-scale cloudlets \citep{mccourt18,liang18,sparre2018physics}. In such scenarios, the cold gas has a characteristic scale $l_{\rm cloudlet} \sim (c_{\rm s} t_{\rm cool})_{\rm c} \sim 0.1 {\rm pc} (n/{\rm cm^{-3}})^{-1}$, where quantities are evaluated in the cold gas phase. The area covering fraction $f_{\rm A}$ (or the average number of cloudlets intercepted along a line of sight) is given by \citep{mccourt18}:
\begin{equation}
f_{\rm A} \sim \frac{D}{l_{\rm cloudlet}} f_{\rm V} 
\end{equation}
where $D$ is the size of the system, $l_{\rm cloudlet}$ is the size of an individual cloudlet, and $f_{\rm V}$ is the volume filling fraction of cold gas. Since ${D}/{l_{\rm cloudlet}} \sim 10^{5}-10^{6}$ on CGM scales, very large covering factors $f_{\rm A} \sim 10^{2}-10^{3}$ are possible despite a small cold gas volume filling fraction $f_{\rm V} \sim 10^{-3}$. The viability of TMLs as an explanation for the observed absorption line systems seen in the CGM remains an open question, but it likely requires the cold gas to be structured on very small scales to have a high area to volume covering fraction.

\section{Acknowledgements}

 We thank Max Gronke and Jim Stone for helpful conversations, and the referee Michael Shull for a constructive and helpful report which improved our paper. This research has used the Extreme Science and Engineering Discovery Environment (XSEDE allocations TG-AST140086). We have made use of NASA's Astrophysics Data System and the yt astrophysics analysis software suite \citep{turk2010yt}. SPO thanks the law offices of May Oh \& Wee for hospitality. PM acknowledges a College of Creative Studies Summer Undergraduate Research Fellowship.

\bibliographystyle{mnras}
\bibliography{master_references}

\end{CJK}
\end{document}